\documentclass[journal,twoside,web]{ieeecolor}

\usepackage{arxiv}
\usepackage{cite}
\usepackage{amsmath,amssymb,amsfonts}
\usepackage{graphicx}
\usepackage{textcomp}

\usepackage{algorithm, algorithmicx, algpseudocode}
\usepackage{subcaption}  
\usepackage{booktabs}        
\usepackage{threeparttablex}  
\usepackage{array}
\usepackage{bm}
\usepackage{bibunits}

\algtext*{EndFor}
\algtext*{EndIf}
\algtext*{EndWhile}
\algtext*{EndSwitch}
\algrenewcommand\alglinenumber[1]{\scriptsize #1~}

\IEEEtitleabstractindextext{

\begin{abstract}

\noindent \textbf{Purpose:} To develop a computationally viable autofocus method for estimating 3D rigid motion in MR imaging.

\vspace{0.25em}

\noindent \textbf{Theory and Methods:} The proposed method, REACT, assumes a piecewise-constant motion trajectory and estimates the rigid motion parameters of individual temporal segments by optimizing an image-quality metric. Coordinate descent is adopted to decompose the high-dimensional optimization problem into a series of subproblems, each updating the motion parameters of a single temporal segment. The cost function of each subproblem is assumed to be approximately locally convex under suitable acquisition conditions. Each subproblem is then solved using a derivative-free solver, thereby avoiding an exhaustive grid search. Numerical simulations investigated the local convexity assumption and data acquisition requirements. REACT was evaluated for respiratory motion correction on \textit{in vivo} free-breathing coronary MR angiography datasets. Coronary artery sharpness was quantified using unbounded image edge profile acutance (u-IEPA).

\vspace{0.25em}

\noindent \textbf{Results:} In numerical simulations, the objective surfaces of the subproblems were approximately locally convex when the current motion estimate was sufficiently close to the desired solution, and REACT required the data collected within each temporal segment to be sufficiently distributed across $k$-space. In the \textit{in vivo} study, REACT yielded higher u-IEPA for both the left anterior descending artery (LAD) and the right coronary artery than did a conventional translational motion-estimation method using image-based navigators. REACT also yielded higher u-IEPA for the LAD than did a conventional autofocus nonrigid motion correction method.

\vspace{0.25em}

\noindent \textbf{Conclusion:} This study demonstrates the feasibility of coordinate descent for autofocus motion correction in MR imaging.

\end{abstract}

}

\begin{document}
\title{Rigid Motion Estimation using Accelerated Coordinate Descent (REACT) for MR Imaging}
\author{
    Kwang Eun Jang and Dwight G. Nishimura
    \thanks{This is a preprint of a manuscript submitted to Magnetic Resonance in Medicine (MRM). Corresponding author: KEJ (kejang@gmail.com).}
}

\maketitle



\section{Introduction}\label{sec:intro}

MR imaging is inherently motion-sensitive because any motion over multiple repetition times (TRs) introduces inconsistencies in the $k$-space data~\cite{zaitsev_motion_2015}. To address motion, numerous methods have been proposed~\cite{zaitsev_motion_2015}. This work falls within the category of autofocus methods~\cite{atkinson_automatic_1997}, also known as autocorrection methods~\cite{mcgee_image_2000}. In these methods, motion is estimated by optimizing a metric that quantifies reconstructed image quality, because motion-induced inconsistencies in $k$-space data typically manifest as image degradations such as blurring and ghosting.

In the autofocus approach, the motion trajectory is commonly modeled as piecewise constant over time, reducing the motion-estimation problem to finding the motion parameters for each temporal segment. Early autofocus methods employed grid searches, often combined with scheduling heuristics such as coarse-to-fine and multiresolution approaches~\cite{atkinson_automatic_1997, mcgee_image_2000, atkinson_automatic_1999}. Subsequently, an analytic derivative of an autofocus metric with respect to rigid motion parameters was derived to enable gradient-based optimization~\cite{loktyushin_blind_2013}. A data-consistency approach was also proposed, which jointly estimates the motion-free image and motion trajectory using a reduced forward model~\cite{haskell_targeted_2018}. The autofocus approach was further extended to nonrigid motion correction by modeling nonrigid motion as localized translations \cite{cheng_nonrigid_2012,ingle_nonrigid_2014}.

One of the main challenges in the autofocus framework is to solve a high-dimensional optimization problem. Even when restricted to 3D rigid motion, the optimization problem involves $6B$ unknowns\textemdash three for translation and three for rotation\textemdash where $B$ denotes the number of temporal segments.

\begin{figure}[tb]
  \centerline{\includegraphics[height=0.58\textheight]{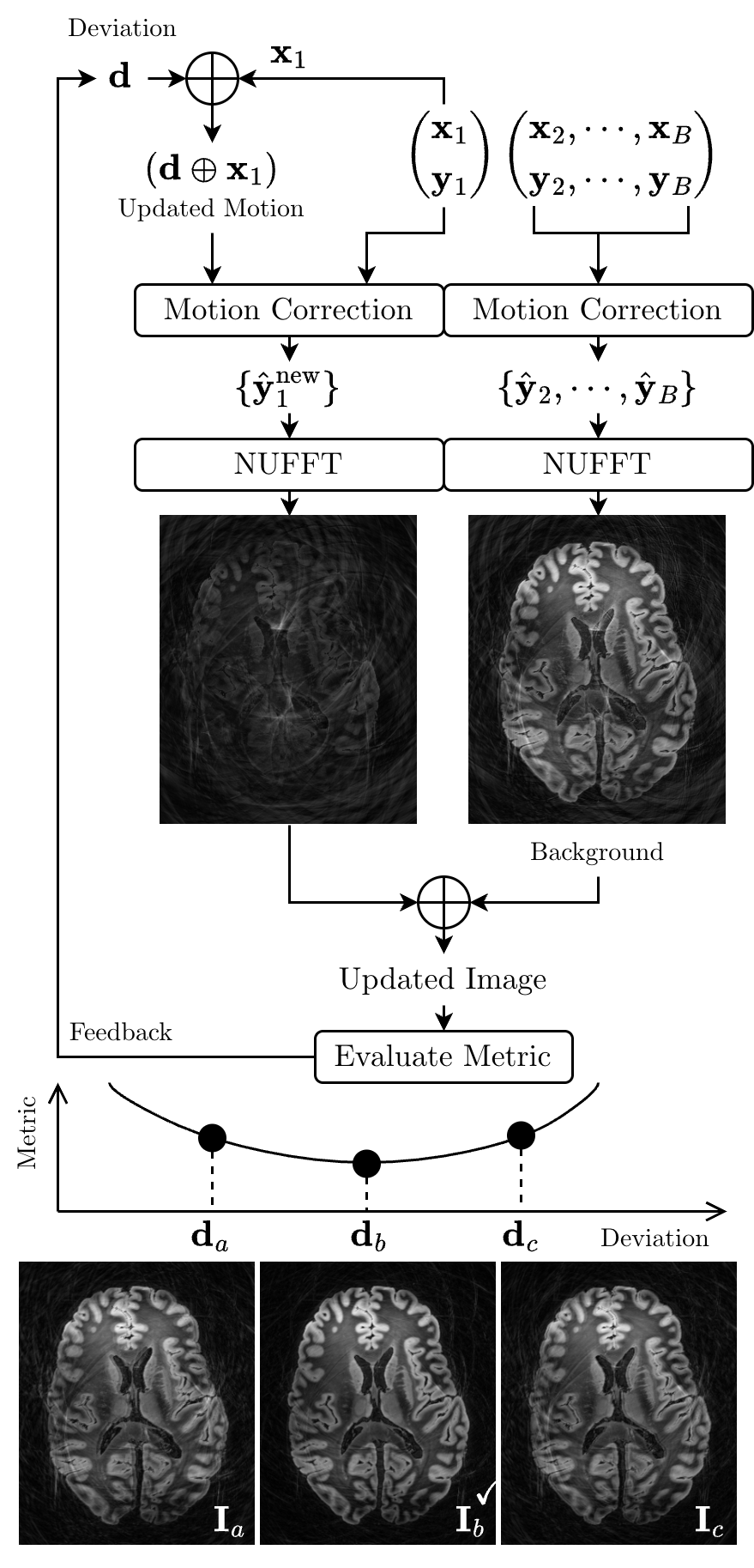}}
  \caption{Conceptual overview of the proposed approach, where the motion parameters of one temporal segment are updated while all others are kept fixed. The cost function of each coordinate-descent subproblem is assumed to be approximately locally convex under suitable acquisition conditions.}
  \label{fig:overview}
\end{figure}

In this work, we adopt the coordinate-descent approach~\cite{sauer_local_1993, bouman_unified_1996, wright_coordinate_2015} for autofocus motion correction \cite{jang_2026_rigid}. The key idea of coordinate descent is to decompose a high-dimensional optimization problem into a series of tractable subproblems, each updating a single \textit{coordinate} while keeping all others fixed. In our setup, each coordinate corresponds to the motion parameters of one temporal segment, and each subproblem therefore involves only six unknowns. A conceptual overview of the proposed framework is shown in Figure~\ref{fig:overview}.

We hypothesize that image quality varies smoothly as the motion parameters of a \textit{single} temporal segment are adjusted, such that the associated cost function is approximately locally convex, provided that the current motion estimate is sufficiently close to the desired solution. This local convexity likely arises because, as this small subset of motion parameters is perturbed, the \textit{active} component (i.e., the Fourier transform of the corresponding $k$-space data) interferes constructively or destructively with the \textit{static} background (i.e., the Fourier transform of the remaining data), thereby altering the levels of blurring and ghosting. The objective functions of the coordinate-descent subproblems may therefore exhibit local convexity around the deviation that yields maximal coherence between the active component and the static background. This hypothesis not only supports the coordinate-descent approach for autofocus motion correction but also eliminates the need for exhaustive grid searches, since local convexity enables modern numerical solvers to efficiently solve the subproblems.

For our coordinate-descent approach, we impose a few conditions on data acquisition. We assume that the $k$-space data collected within each temporal segment (a) are sufficiently distributed across $k$-space and (b) have nontrivial energy in the image domain. The former enables motion estimation along all degrees of freedom, and the latter ensures that perturbing the motion parameters for any single temporal segment produces measurable changes in an autofocus metric. Segmented acquisitions can satisfy these conditions if the sampling trajectory within each acquisition window provides broad $k$-space coverage. We also assume that the signal-to-noise ratio (SNR) is adequate.

Coordinate-wise updates, however, may converge slowly or yield a suboptimal solution when low-frequency discrepancies are present between the current estimates and the desired solution. In the context of motion estimation, this issue can arise when periodic motion induces correlated (low-frequency) errors across groups of temporal segments. In addition, without good initialization, the coordinate-descent approach may yield inaccurate motion estimates because the assumed local convexity of the coordinate-wise cost function may no longer hold.

To address these issues, we introduce Rigid motion Estimation using Accelerated Coordinate descenT (REACT). The temporal segments are partitioned into progressively smaller groups over the iterations, with each subproblem corresponding to a group rather than an individual segment. We show that applying these group-wise updates in the early iterations can improve convergence when low-frequency errors are present in the initial motion estimates. By reducing the number of subproblems, this strategy also reduces computation time. We present successive 1D sweeps and a learning-rate ramp-up to improve the robustness of REACT when good initial estimates are unavailable.

This study primarily describes and evaluates REACT for free-breathing coronary MR angiography. The proposed framework could be adapted to other MR applications by tailoring the partitioning strategy to the expected motion characteristics and available auxiliary data.

A Python implementation of REACT will be made publicly available on GitHub upon publication of this work. 

\section{Theory}\label{sec:theory}

Table~\ref{tab:notation} summarizes the notation used in this paper. Following prior work~\cite{atkinson_automatic_1997, mcgee_image_2000, atkinson_automatic_1999, loktyushin_blind_2013}, the motion trajectory is assumed to be piecewise constant in time over the course of the segmented acquisition. In this paper, a ``beat'' refers to the smallest time interval during which the object is assumed to remain stationary (i.e., one temporal segment). In free-breathing coronary MRA, for example, the heart is typically assumed to remain stationary during the relatively short acquisition window of every heartbeat~\cite{bernstein2004handbook}.

\begin{table}[tb] %
\caption{Notation.\label{tab:notation}}
\vspace{6pt}
{%
\renewcommand{\arraystretch}{1.08}
\begin{tabular*}{\columnwidth}{@{}%
  >{\raggedright\arraybackslash}p{0.16\columnwidth}%
  @{}%
  >{\raggedright\arraybackslash}p{0.66\columnwidth}%
  @{}%
  >{\centering\arraybackslash}p{0.17\columnwidth}%
@{}}
\toprule
\textbf{Symbol} & \textbf{Description} & \textbf{Size}\\
\midrule
$B$ & Number of temporal segments (beats) & $(1 \times 1)$ \\
$b$ & Beat index & $(1 \times 1)$ \\
$N$ & Number of iterations & $(1 \times 1)$ \\
$n$ & Iteration index & $(1 \times 1)$ \\
$M$ & Number of partitions & $(1 \times 1)$ \\
$m$ & Partition index & $(1 \times 1)$ \\
$q$ & Group index within a partition & $(1 \times 1)$ \\
$\square_b$ & Quantity at the $b$-th beat & N/A \\
$\square^{(n)}$ & Quantity at iteration $n$ & N/A \\
$\square^{(-1)}$ & Uncorrected data or trajectory & N/A \\
$t$ & Time & $(1 \times 1)$ \\
$\mathbf{k}(t)$ & Sampling trajectory at $t$ & $(3 \times 1)$ \\
$\mathbf{s}(t)$ & Density-corrected $k$-space data at $t$ & $(1 \times 1)$ \\
$\mathbf{t}$ & Translation vector & $(3 \times 1)$ \\
$\mathbf{v}$ & Rotation vector (rotvec) & $(3 \times 1)$ \\
$\mathbf{x}$ & Rigid vector $(\mathbf{x}^T = [\mathbf{t}^T, \mathbf{v}^T])$& $(6 \times 1)$ \\
$\mathbf{X}$ & Stacked rigid vectors (i.e., $[\mathbf{x}_1,\!\cdots,\!\mathbf{x}_B])$& $(6 \times B)$ \\
$\oplus$ & Addition operator for rigid vectors & N/A \\
$\mathbf{R}(\mathbf{v})$ & Rotation matrix induced by $\mathbf{v}$ & $(3 \times 3)$ \\
$\hat{\mathbf{d}}$ & Optimal deviation of a rigid vector & $(6 \times 1)$ \\
$\mathbf{d}_{m,b}$ & Deviation for the $b$-th beat in partition $m$ & $(6 \times 1)$ \\
$\mathbf{G}_m$ & $m$-th partition of beat indices & N/A \\
$\mathbf{g}_{m, q}$ & $q$-th group in partition $m$ & N/A \\
$|\mathbf{g}_{m, q}|$ & Number of beats in group $\mathbf{g}_{m, q}$ & $(1 \times 1)$ \\
$|\mathbf{G}_m|$ & Number of groups in partition $\mathbf{G}_m$ & $(1 \times 1)$ \\
$f(\mathbf{x}_b|\mathbf{X})$ & Cost function with respect to $\mathbf{x}_b$ while the other rigid vectors are fixed. & N/A \\
$\alpha^{(n)}$ & Learning rate at iteration $n$ & $(1 \times 1)$ \\
\bottomrule
\end{tabular*}
}%
\end{table}

\subsection{Rigid Motion Correction in k-space}\label{subsec:theory-rigid_in_ksp}

We use rotation vectors (rotvecs) $\mathbf{v}$ to represent 3D rotations. A rotvec is a 3D vector whose direction and magnitude indicate the rotation axis and rotation angle (in radians), respectively~\cite{lynch2017modern}. The corresponding rotation matrix $\mathbf{R}(\mathbf{v})$ is synthesized using Rodrigues' formula~\cite{lynch2017modern}.

Rigid motion in image space can be represented in $k$-space through the properties of the Fourier transform~\cite{bracewell2012fourier}. Specifically, a translation in image space introduces a linear phase in $k$-space, while a rotation in image space results in a rotation of the sampling trajectory. Our goal is to estimate the translation vector $\mathbf{t}_b$ and rotation vector $\mathbf{v}_b$ for each beat $b$, such that the corresponding sampling trajectory and $k$-space data are corrected as
\begin{align}
	\mathbf{k}_b(t) &= \mathbf{R}(\mathbf{v}_b)\cdot\mathbf{k}_b^{(-1)}(t),\label{eq:traj_corr_ksp}\\
	\mathbf{s}_b(t) &= \mathbf{s}_b^{(-1)}(t) \exp \left(j 2\pi \left(\mathbf{t}_b^T \cdot \mathbf{k}_b^{(-1)}(t) \right)\right),\label{eq:rigid_in_ksp}
\end{align}
\noindent where $\mathbf{k}_b^{(-1)}(t)$ and $\mathbf{s}_b^{(-1)}(t)$ denote the uncorrected sampling trajectory and $k$-space data of the $b$-th beat, respectively.

\subsection{Image Gradient Entropy}\label{subsec:theory-entropy}

We use image gradient entropy as the image-quality metric to be minimized, following the recommendation from the comparative study of McGee \textit{et~al.}~\cite{mcgee_image_2000} on metrics for autocorrection approaches. Because the motion correction described by \eqref{eq:traj_corr_ksp} and \eqref{eq:rigid_in_ksp} is applied globally to the entire image, we introduce a binary mask $\mathbf{W}$ to localize the metric to a region of interest (ROI), such as the heart. The image gradient entropy $h$ used in this work is computed as
\begin{align}
    h &= -\sum_{\text{voxel}} \overline{\mathbf{H}} \log_2(\overline{\mathbf{H}}),\label{eq:entropy}\\
    \overline{\mathbf{H}} &= \frac{\mathbf{H}}{\sum_{\text{voxel}} \mathbf{H}},\\
    \mathbf{H} &= \mathbf{W} \sqrt{\left| \nabla_x \mathbf{I} \right|^2 + \left| \nabla_y \mathbf{I} \right|^2 + \left| \nabla_z \mathbf{I} \right|^2}\label{eq:abs_gradient},
\end{align}
\noindent where $\mathbf{I}$ is the reconstructed image, $\mathbf{H}$ is the masked image gradient magnitude, and $\nabla_x$, $\nabla_y$, and $\nabla_z$ are finite-difference operators along the $x$, $y$, and $z$ directions, respectively. Unlike the separable entropy used by Loktyushin et al.~\cite{loktyushin_blind_2013}, which is defined as the sum of the entropies of the individual gradient components, the metric in \eqref{eq:entropy} is rotationally invariant.

\subsection{Coordinate Descent}\label{subsec:theory-icd}

The key features of the coordinate-descent approach are: (a) updating one \textit{coordinate} at a time while keeping the others fixed, (b) immediately using the updated coordinate in subsequent updates, and (c) updating the coordinates in a randomized order to prevent systematic bias.

Algorithm~\ref{alg:CD} presents the pseudocode for a direct adaptation of coordinate descent to the motion-estimation problem. For each beat, a numerical solver estimates the optimal deviation from the current motion estimate, and this process is repeated across all temporal segments in a randomized order. Based on the hypothesis that each coordinate-wise subproblem has an approximately locally convex cost function, we employ the numerical solver to update the motion parameters rather than performing exhaustive grid searches. In this work, we choose a derivative-free solver, which does not require analytic derivatives of the image-quality metric in \eqref{eq:entropy} with respect to the motion parameters.


\begin{algorithm}[tb]
\caption{Direct Adaptation of Coordinate Descent}\label{alg:CD}
\begin{algorithmic}[1]
  \State Initialize:
  \State $\quad\mathbf{X}^{(0)} \leftarrow \left(\mathbf{x}_1^{(0)}, \cdots, \mathbf{x}_B^{(0)}\right)$
  \For{$n=\left(1, \cdots, N\right)$}
    \State $\mathbf{X}^{(n)} \leftarrow \mathbf{X}^{(n-1)}$
    \State $\mathbf{g} \leftarrow \text{randomize}~\{1, \cdots, B\}$
    \For{$b \in \mathbf{g}$}
      \State $\hat{\mathbf{d}} \leftarrow \text{argmin}_{\mathbf{d}}~f\left(\mathbf{d} \oplus \mathbf{x}_b^{(n-1)} \mid \mathbf{X}^{(n)}\right)$
      \State $\mathbf{x}_b^{(n)} \leftarrow \alpha^{(n)}\hat{\mathbf{d}} \oplus \mathbf{x}_b^{(n-1)}$
    \EndFor
  \EndFor
\end{algorithmic}
\vspace{-0.9\baselineskip}\par\noindent\hrulefill
\par\footnotesize {
Line 7: Using $\mathbf{X}^{(n)}$ reflects immediate use of the updated coordinate.
}
\end{algorithm}

\subsection{Group-Wise Update}\label{subsec:theory-group}


To improve convergence in the presence of low-frequency errors, we incorporate group-wise updates in the early iterations. This strategy can also reduce computation time by reducing the number of subproblems.

Algorithm~\ref{alg:REACT} presents the pseudocode for the proposed REACT method. At each iteration, the temporal segments are partitioned into groups whose beats are expected to have similar motion-estimation errors. A single optimal deviation is then estimated for each group and applied to every beat in that group. The group sizes are progressively reduced over the iterations; when every group is a singleton ($\mathbf{g}_{q} = \{q\}$), REACT becomes equivalent to Algorithm~\ref{alg:CD}.

To reduce sensitivity to group boundaries, Algorithm~\ref{alg:REACT} uses multiple partitions. It first computes optimal deviations independently for each partition. Each beat is then updated using the mean of its deviations across partitions.

The partitioning method must be tailored to the specific application and the available prior information or auxiliary data. In free-breathing coronary MRA, respiratory motion across different heartbeats tends to be similar when those heartbeats occur at the same respiratory phase. We therefore assume that beats with similar motion-estimation errors can be identified by classifying the temporal segments into groups. In our study, an image-based navigator (iNAV) is acquired at each heartbeat, and spatial features extracted from the iNAVs serve as feature vectors for clustering. External physiological signals, such as respiratory traces, can also be incorporated. In brain MRI, motion often drifts slowly over time~\cite{maclaren_measurement_2012,schilling_head_2025}; therefore, the temporal segments can be partitioned into contiguous groups, allowing REACT to estimate motion from coarser to finer temporal scales.

\begin{algorithm}[t]
\caption{Pseudocode for REACT}\label{alg:REACT}
\begin{algorithmic}[1]
  \State Initialize:
  \State $\quad\mathbf{X}^{(0)} \leftarrow \left(\mathbf{x}_1^{(0)}, \cdots, \mathbf{x}_B^{(0)}\right)$
  \For{$n=\left(1, \cdots, N\right)$}
    \State $\{\mathbf{G}_m\}_{m=1}^{M} \leftarrow \text{generate } M~\text{partitions of}~ \{1,\!\cdots\!,B\}$
    \For{$m=\left(1, \cdots, M\right)$}
      \State $\widetilde{\mathbf{X}} \leftarrow \mathbf{X}^{(n-1)}$
      \For{$q=(1, \ldots, |\mathbf{G}_m|)$}
        \State $\hat{\mathbf{d}} \leftarrow \text{argmin}_{\mathbf{d}}~f\left(\mathbf{d} \oplus \mathbf{X}^{(n-1)}[\mathbf{g}_{m,q}] \mid \widetilde{\mathbf{X}}\right)$
        \For{$b \in \mathbf{g}_{m,q}$}
          \State $\mathbf{d}_{m,b} \leftarrow \alpha^{(n)}\hat{\mathbf{d}}$
          \State $\widetilde{\mathbf{x}}_{b} \leftarrow \mathbf{d}_{m,b} \oplus \mathbf{x}^{(n-1)}_{b}$
        \EndFor
      \EndFor
    \EndFor
    \For{$b=(1, \cdots, B)$}
      \State $\overline{\mathbf{d}}_b \leftarrow \dfrac{1}{M}\sum_{m=1}^{M}\mathbf{d}_{m,b}$
      \State $\mathbf{x}_b^{(n)} \leftarrow \overline{\mathbf{d}}_b \oplus \mathbf{x}_b^{(n-1)}$
    \EndFor
  \EndFor
\end{algorithmic}
\vspace{-0.9\baselineskip}\par\noindent\hrulefill
\par\footnotesize {
Line 6: The deviations for each partition are computed independently from the same starting state.
}
\par\footnotesize {
Lines 8--11: A single optimal deviation is computed for a group and applied to every beat in that group.
}
\par\footnotesize {
Line 13: The partition-specific deviations are averaged to reduce sensitivity to group boundaries.
}
\end{algorithm}

\subsection{Optional 1D Sweeps and Learning-Rate Ramp-Up}\label{subsec:theory-sweep_and_ramp}

When good initial estimates are unavailable, REACT may yield inaccurate estimates because the coordinate-wise cost function may no longer exhibit local convexity. To improve robustness in such scenarios, we optionally perform a coarse grid search on one temporal segment before each coordinate-wise optimization to provide an initial guess for translational motion. Instead of a full 3D search, we use three successive 1D sweeps, since the additional phase term due to translation can be factorized into separable components along each spatial direction. When motion exhibits a dominant direction\textemdash such as the superior--inferior (SI) direction in respiratory motion\textemdash this strategy can be improved by first sweeping along the dominant direction.

To further improve stability, we optionally use smaller learning rates $\alpha^{(n)}$ during early iterations. This is because conservative early updates help avoid overly aggressive motion corrections before a coherent background has formed (i.e., before the objective function becomes approximately locally convex). We use a simple linear ramp-up schedule for $\alpha^{(n)}$.

\subsection{Self-Navigating REACT (SN-REACT)}\label{subsec:theory-sn}

Algorithm~\ref{alg:SN-REACT} presents the pseudocode for self-navigating REACT (SN-REACT), which relies solely on imaging data. At each outer iteration, a new sequence of partitions---one for each inner iteration---is constructed from the current motion estimates and the deviations from the preceding REACT update. Group sizes progressively decrease across the inner iterations, so that the final partition consists of singleton groups. SN-REACT can therefore be viewed as a series of alternating group-wise and coordinate-wise updates: the group-wise updates gradually reduce low-frequency discrepancies, while the deviations from the final coordinate-wise update guide the construction of partitions for the next outer iteration. In the initial step, a single REACT update with singleton groups and an unattenuated learning rate ($\alpha\!=\!1$) is performed solely to obtain a self-derived feature for partitioning; the resulting motion estimate is discarded. 

\begin{algorithm}[t]
\caption{Pseudocode for SN-REACT}\label{alg:SN-REACT}
\begin{algorithmic}[1]
  \State Initialize:
  \State $\quad\mathbf{g} \leftarrow \text{randomize}~\{1, \cdots, B\}$
  \State $\quad\mathbf{X}^{(0)} \leftarrow \mathbf{0}$
  \State $\quad\left(\underline{\phantom{\mathbf{X}}}, \mathbf{D}^{(0)}\right) \leftarrow \operatorname{REACT}\left(\vphantom{\mathbf{G}_m^{(r)}}\mathbf{X}^{(0)}; \{\mathbf{g}\}, \alpha\!=\!1\right)$
  \For{$n=\left(1, \cdots, N\right)$}
    \State $\bigl\{\{\mathbf{G}_m^{(r)}\}\bigr\}_{r=1}^{R}\leftarrow \operatorname{partition}\left(\vphantom{\mathbf{G}_m^{(r)}}\mathbf{X}^{(n-1)},\mathbf{D}^{(n-1)}\right)$
    \State $\mathbf{X}^{(n, 0)} \leftarrow \mathbf{X}^{(n-1)}$
    \For{$r=(1, \cdots, R)$}
      \State $\left(\mathbf{X}^{(n, r)}, \mathbf{D}^{(n, r)}\right) \leftarrow$
      \Statex $\quad\quad\quad\quad\operatorname{REACT}\left(\mathbf{X}^{(n, r-1)}; \{\mathbf{G}_m^{(r)}\}, \alpha\!=\!\alpha^{(n)}\right)$
    \EndFor
    \State $\mathbf{X}^{(n)}, \mathbf{D}^{(n)} \leftarrow \mathbf{X}^{(n, R)}, \mathbf{D}^{(n, R)}$
  \EndFor
\end{algorithmic}
\vspace{-0.9\baselineskip}\par\noindent\hrulefill
\par\footnotesize {
$\operatorname{REACT}(\cdot)$ denotes one REACT update for given partitions and a learning rate. $\mathbf{D}^{(n)}$ contains per-beat deviations at outer iteration $n$.
}
\par\footnotesize {
Line 4: A single REACT update using singleton groups and the unattenuated learning rate ($\alpha\!=\!1$) bootstraps the grouping feature.
}
\par\footnotesize {
Line 6: Based on the current motion estimates and preceding REACT deviations, a new sequence of partitions is constructed, with group sizes progressively decreasing to singletons.
}
\end{algorithm}

\section{Method}\label{sec:method}

Numerical simulations using 2D motion models were conducted to (a) examine the local convexity of coordinate-wise cost functions, (b) evaluate REACT with different sampling trajectories at varying levels of initial estimation error, and (c) compare the performance of REACT with and without group-wise updates in the presence of low-frequency errors. REACT was then evaluated for respiratory motion correction using \textit{in vivo} free-breathing, cardiac-gated coronary MR angiography datasets.

\subsection{Numerical Simulations}\label{subsec:method-simul}

A numerical brain phantom was generated by downsampling an \textit{ex vivo} brain MR image~\cite{edlow_7_2019} from 100~\textmu m to 0.5~mm and zero-padding it to a matrix size of $384 \times 320$. Complex Gaussian noise was added to the $k$-space data to obtain an SNR of 8.0 in the sum-of-squares (SoS) reconstruction, measured within an elliptical ROI enclosing the brain.

Two 2D acquisition schemes were simulated: interleaved spiral imaging and interleaved multi-shot echo-planar imaging (EPI). The spiral trajectories were designed using parameters similar to those described by Kasper et al.~\cite{kasper_rapid_2018}. For simplicity, the interleaved EPI trajectories were constructed by decimating the $k_y$ lines of a 2D Cartesian grid.

Two motion models were considered: uncorrelated and correlated motion. The uncorrelated motion was drawn from a zero-mean uniform distribution, providing a favorable setting for coordinate-wise updates. The correlated motion was generated using a Gaussian random walk and was scaled to a target displacement rate of 9~mm/min. For reference, Reuter et al. reported average root-mean-square displacement rates of approximately 3.3, 5.3, and 12.5~mm/min for scans classified as ``pass'', ``warn'', and ``fail'', respectively, in a manual quality-control assessment~\cite{reuter_head_2015}. The range of rotational motion was set to half that of translational motion to simulate translationally dominant motion.

For group-wise updates, the temporal segments were initially divided into three contiguous groups. The number of groups was then doubled at each iteration, approximately halving the group size. At each iteration, three partitions ($M\!=\!3$) with different starting offsets were used to soften group boundaries.

The Supporting Information provides the REACT parameters and detailed numerical simulation settings, including SNR selection, acquisition parameters, and the procedure used to generate the simulated motion trajectories.

\subsection{\textit{In Vivo} Data Acquisition}\label{subsec:method-data}

Ten \textit{in vivo} datasets were acquired on a GE Signa 1.5~T scanner (GE Healthcare, Waukesha, WI) using an 8-channel cardiac array coil. Data were collected over 610 heartbeats (7--10 minutes total) while eight healthy volunteers (1F, 7M) breathed normally. The study was approved by the Institutional Review Board, and written informed consent was obtained from all subjects. Eight datasets were acquired using leading sagittal and trailing coronal 2D iNAVs~\cite{wu2013free} (resolution = 3.1~mm), and two datasets were acquired using trailing 3D iNAVs~\cite{addy20173d} (resolution = 4.4~mm). A 3D cones phyllotaxis trajectory~\cite{malave_whole-heart_2019} was employed, acquiring 18 conic interleaves per heartbeat (temporal resolution = 98~ms). The prescribed FOV and spatial resolution were 28~cm~$\times$~28~cm~$\times$~14~cm and 1.2~mm~$\times$~1.2~mm~$\times$~1.25~mm, respectively~\cite{malave_whole-heart_2019}. An alternating repetition time balanced steady-state free precession (ATR-bSSFP) sequence~\cite{leupold2006alternating} was used with the following parameters~\cite{wu2013free, malave_whole-heart_2019}: TR = 5.4~ms, TE = 0.6~ms, FA = 70$^\circ$, and BW = $\pm$125~kHz.

\subsection{Motion-Estimation Methods Compared}\label{subsec:method-compared}



The following abbreviations are used for the approaches compared in the \textit{in vivo} study. \texttt{NoCo} denotes no motion correction and serves as the baseline. \texttt{Tran} represents conventional translational motion estimation using iNAVs. \texttt{REACT} refers to Algorithm~\ref{alg:REACT}, initialized with the \texttt{Tran} estimates. \texttt{SN-REACT} refers to Algorithm~\ref{alg:SN-REACT}, which relies solely on imaging data (i.e., without iNAVs). \texttt{REACT-Tran} assumes translational motion only. \texttt{REACT-NoGrp} and \texttt{SN-REACT-NoGrp} denote \texttt{REACT} and \texttt{SN-REACT} without group-wise updates, respectively.

For additional comparison, the autofocus nonrigid motion correction method proposed by Ingle et al.~\cite{ingle_nonrigid_2014} was also implemented. This method was selected as a comparator alongside \texttt{Tran} because it is an autofocus approach developed for free-breathing coronary MRA using a 3D cones trajectory, closely matching the acquisition setting of this study. In this approach, a motion basis set is defined to generate a finite set of motion-corrected images~\cite{cheng_nonrigid_2012,ingle_nonrigid_2014}. The final motion-corrected reconstruction is then assembled from this set by selecting, at each voxel location, the voxel value from the image with the lowest local image gradient entropy~\cite{cheng_nonrigid_2012,ingle_nonrigid_2014}. As in the method of Ingle et al.~\cite{ingle_nonrigid_2014}, the motion basis sets were constructed by scaling the estimated translational motion trajectories. \texttt{NR-Tran} and \texttt{NR-REACT} denote the nonrigid motion corrections based on the estimated translational motion trajectories obtained from \texttt{Tran} and \texttt{REACT}, respectively.

\subsection{Implementation}\label{subsec:method-implementation}


The Bound Optimization BY Quadratic Approximation (BOBYQA) algorithm~\cite{powell2009bobyqa}, implemented in the NLopt package~\cite{johnson_nlopt} (\verb|LN_BOBYQA|), was used as the numerical solver. 

For the group-wise update in REACT, spatial features were extracted from the iNAVs using principal component analysis (PCA). The temporal segments were then classified into multiple respiratory phases using recursive k-means clustering. The number of clusters was initially set to 3 and progressively increased over four iterations, approximately halving the size of the largest cluster at each iteration. In the fifth iteration, the coordinate-wise update was applied. Further details of the grouping procedures and parameter settings for \texttt{REACT} and \texttt{SN-REACT} are provided in the Supporting Information.



During REACT, an accelerated SoS method was used. The reconstruction for each coil was decomposed into a dynamic component\textemdash gridded from the subset of data being updated\textemdash and a static background reconstructed from the remaining data. This implementation significantly reduced the overall computation time by allowing the static background to be precomputed while gridding only a small portion of the data on the fly. 



Simple 3D ellipsoids with manually selected centers and semi-axes were used as binary ROI masks ($\mathbf{W}$ in \eqref{eq:abs_gradient}) to compute the image-quality metric; these parameter choices may affect motion estimation. 

The REACT methods were run on Amazon Elastic Compute Cloud (Amazon EC2) using g5.xlarge instances, each equipped with 4 vCPUs, 16~GiB of RAM, and one NVIDIA A10G GPU.

Implementation details for the initial translational motion estimates (\texttt{Tran}), the baseline autofocus nonrigid motion-correction methods (\texttt{NR-Tran} and \texttt{NR-REACT})~\cite{ingle_nonrigid_2014}, and the final motion-corrected image reconstruction are provided in the Supporting Information.

\subsection{Assessment}\label{subsec:method-assessment}
For a quantitative assessment of coronary artery sharpness, we adapt the image edge profile acutance (IEPA) metric~\cite{biasiolli_loss_2011} as follows:
\begin{align}
    \text{u-IEPA} &= \frac{\text{Mean}(\{a_j\})}{\text{Max}(\{\mathbf{r}_j\}) - \text{Min}(\{\mathbf{r}_j\})},\\
    a_j &= \text{RMS}\left(\left\{\frac{\mathbf{r}_j[i+1] - \mathbf{r}_j[i]}{\Delta}\right\}\right),
\end{align}
\noindent where $\mathbf{r}_j$ denotes the $j$-th cross-sectional intensity profile of the coronary artery, and $\Delta$ denotes its spatial sampling interval. Here, ``u'' stands for ``unbounded''; because u-IEPA is normalized by the spatial resolution, it is not bounded within $(0, 1)$ as in the original IEPA. This modification removes the dependency on spatial resolution, facilitating fair comparisons across different setups.

A non-radiologist manually annotated the left anterior descending (LAD) and right coronary (RCA) arteries using the 3D curved multi-planar reformatting (MPR) tool in Horos~\cite{horos}. The centerline paths of the coronary arteries were then generated by in-house software based on the manual annotations. Cross-sectional planes orthogonal to the artery centerline were identified at 1.2~mm intervals along each \textit{curved path}, and 60 radial profiles were extracted from each plane with a spatial sampling interval of 1.2~mm to measure u-IEPA scores. To obtain curved MPR images for visualization, the computationally refined curved paths were exported back to Horos.

\begin{figure*}[tb]
  \centerline{\includegraphics[width=\linewidth]{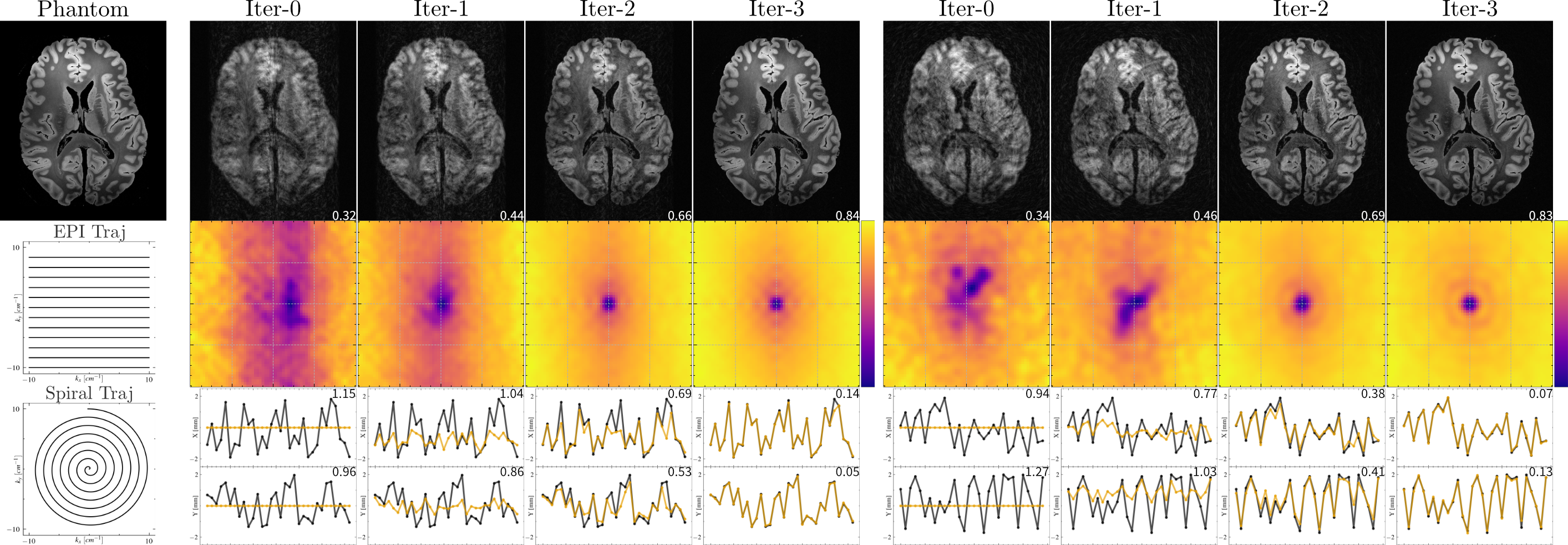}}
  \caption{Numerical simulation (2D translational motion) with EPI and spiral acquisitions. Left panel: numerical phantom and example trajectories for an EPI shot and a spiral interleaf. Center and right panels: results for the EPI and spiral acquisitions, respectively. Top row: motion-corrected reconstructions from iteration~0 (no correction) to iteration~3; values denote structural similarity (SSIM) indices. Middle row: averaged objective values sampled within $\pm 5$~mm of the ground truth. Bottom row: ground-truth (black) and estimated (orange) translational motion; values denote mean root-mean-square errors (RMSEs). As iterations progressed, the objective surfaces near the ground truth increasingly resembled locally convex functions. The objective surfaces for the EPI acquisition exhibited greater anisotropy between the $x$- and $y$-directions than those for the spiral acquisition.}
  \label{fig:simul_cost}
\end{figure*}

\begin{figure*}[tb]
  \centerline{\includegraphics[width=0.975\linewidth]{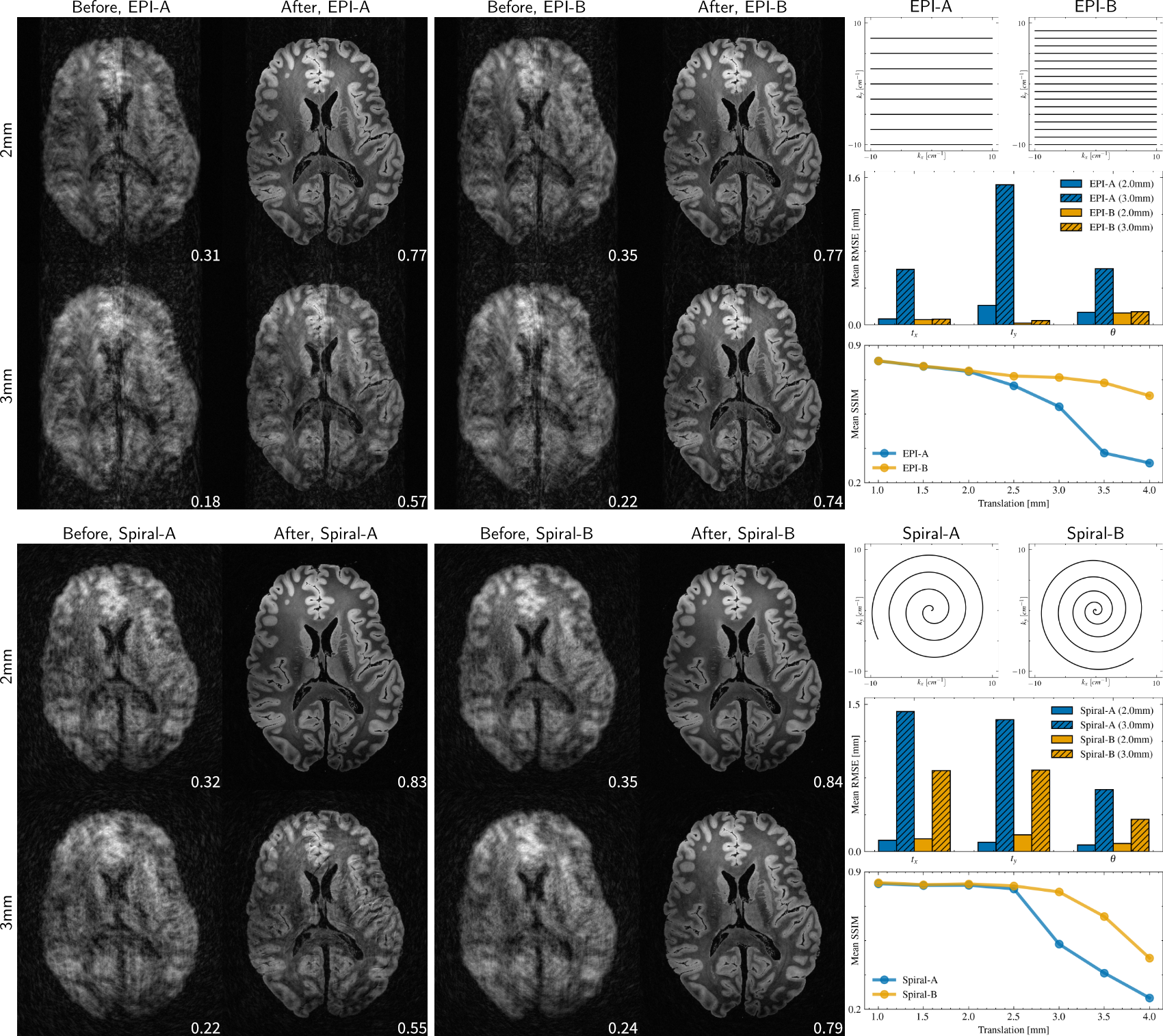}}
  \caption{Numerical simulations (2D rigid motion) comparing REACT across sampling trajectories and translation ranges for EPI (top panel) and spiral (bottom panel) acquisitions. EPI-A and EPI-B used 8 and 16 $k_y$-lines per shot, respectively. Spiral-A used a readout duration of 10~ms, whereas Spiral-B used a readout duration of 20~ms with variable-density sampling ($\alpha = 1.5$). Left panels: representative reconstructions selected from 25 runs, shown before and after motion correction for translation ranges of 2~mm (upper row) and 3~mm (lower row); values denote structural similarity (SSIM) indices. Right panels: representative sampling trajectories (top), mean root-mean-square errors (RMSEs) at translation ranges of 2 and 3~mm (middle), and mean SSIM indices across different motion amplitudes (bottom). For spiral imaging, RMSEs were comparable across motion components after accounting for their prescribed ranges, whereas EPI-A yielded substantially lower RMSEs for $x$-translation than for the other components. At larger motion amplitudes, acquisitions providing denser $k$-space coverage within each temporal segment yielded lower RMSEs and higher mean SSIM.}\label{fig:simul_traj}
\end{figure*}

\begin{figure*}[tb]
  \centerline{\includegraphics[width=0.975\linewidth]{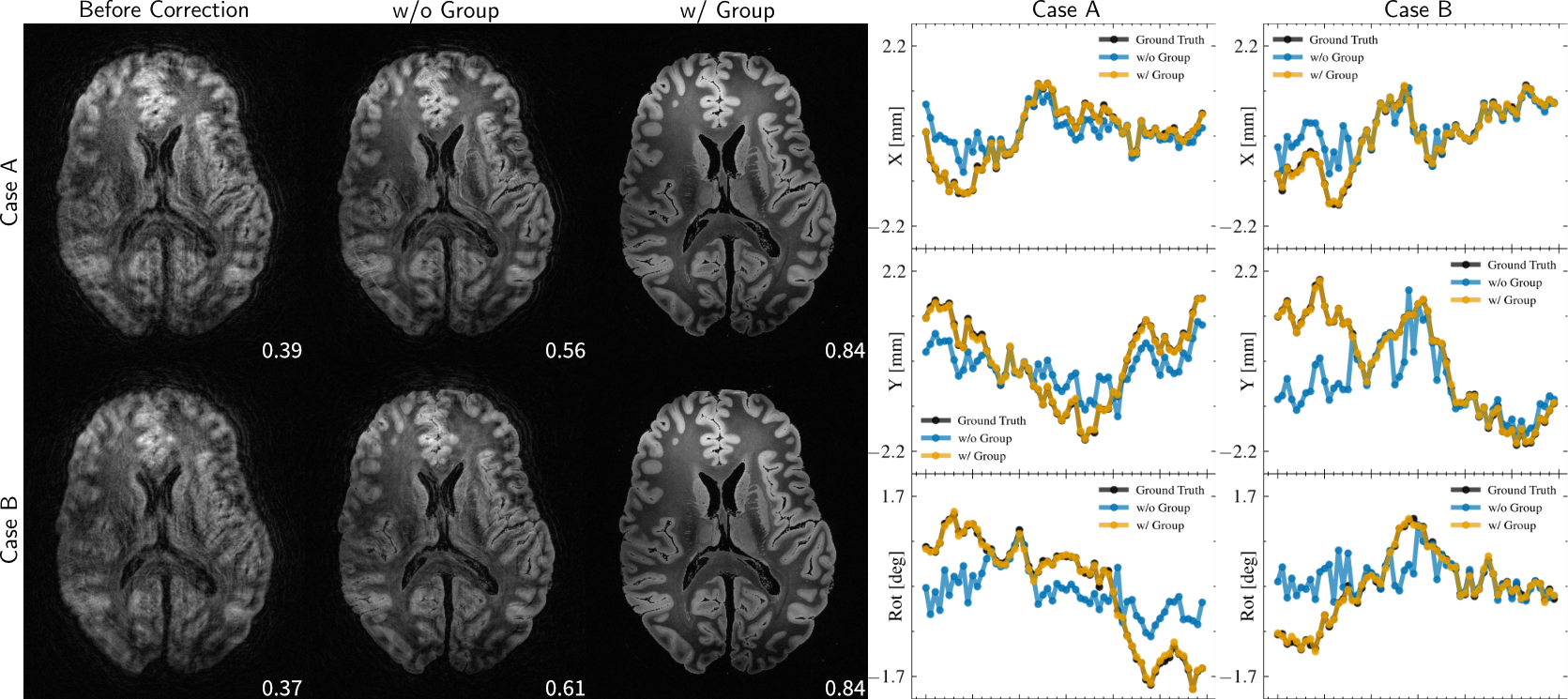}}
  \caption{Numerical simulations (2D rigid motion) comparing REACT with and without group-wise updates for two slowly drifting motion cases. Left panel: reconstructions for Case~A (top) and Case~B (bottom) before motion correction (leftmost) and after motion correction using REACT without (middle) and with (rightmost) group-wise updates; values denote structural similarity (SSIM) indices. Right panel: ground-truth (black) and estimated motion trajectories without (blue) and with (orange) group-wise updates. REACT with group-wise updates accurately captured both the low-frequency drift and higher-frequency motion variations.}\label{fig:simul_group}
\end{figure*}

\section{Results}\label{sec:results}

\subsection{Numerical Simulations}\label{subsec:results-simulation}

Figure~\ref{fig:simul_cost} shows the estimated translational motion, motion-corrected reconstructions, and the corresponding objective values near the ground truth at each REACT iteration. As the REACT iterations progressed, the motion estimates became more accurate, and the objective surface near the ground truth increasingly resembled a locally convex function. These observations are consistent with our hypothesis that, under appropriate acquisition conditions, the cost function of each coordinate-descent subproblem is approximately locally convex when the current motion estimate is sufficiently close to the desired solution.

Figure~\ref{fig:simul_traj} compares the performance of REACT across different sampling trajectories and motion ranges. Because the initial motion estimates were set to zero, a wider motion range corresponds to larger discrepancies between the initial estimates and the ground truth. For smaller initial estimation errors, REACT yielded accurate motion estimates with all sampling trajectories. As the initial estimation errors increased, however, acquisitions providing denser $k$-space coverage within each temporal segment yielded superior results. For spiral imaging, the root-mean-square errors (RMSEs) of the three motion components were comparable after accounting for their different prescribed ranges,\footnote{The simulated rotation range was half the translation range to represent translationally dominant motion.} whereas EPI-A produced substantially lower errors in $x$-translation than in the other motion components. A similar directional difference appears in the objective surfaces shown in Figure~\ref{fig:simul_cost}, with greater anisotropy between the $x$- and $y$-directions for EPI (the readout and phase-encoding directions, respectively) than for spiral imaging.

These findings show that REACT requires the data collected within each temporal segment to be sufficiently distributed across $k$-space. For motion estimation along all degrees of freedom, the sampling trajectory should cover $k$-space in all directions. Dense sampling near the center of $k$-space would ensure that the $k$-space data collected during each temporal segment have nontrivial energy in the image domain, in addition to providing inherent robustness to motion and flow~\cite{glover_projection_1992,meyer_fast_1992,pipe_motion_1999}. Note that these findings are generally consistent with the observations of Cordero-Grande et al. that distributing samples across $k$-space within individual acquisition segments can improve motion estimation~\cite{cordero2016sensitivity,cordero2020motion}.

These findings also indicate that the performance of REACT depends on multiple factors, including the sampling distribution and the accuracy of the initial motion estimates, making it difficult to derive a general theoretical condition for reliable motion estimation.

Figure~\ref{fig:simul_group} compares the performance of REACT with and without group-wise updates for slowly drifting motion. When only coordinate-wise updates were used, REACT converged to suboptimal solutions: it estimated the short-term variations relatively well but failed to capture the low-frequency drift. With group-wise updates, the temporal segments were partitioned into contiguous groups of progressively decreasing size, allowing motion to be estimated from coarser to finer temporal scales. This strategy accurately recovered both the low-frequency drift and higher-frequency motion variations.

\subsection{\textit{In Vivo} Experiments}\label{subsec:results-invivo}


\begin{figure*}[tb]
  \centerline{\includegraphics[width=\linewidth]{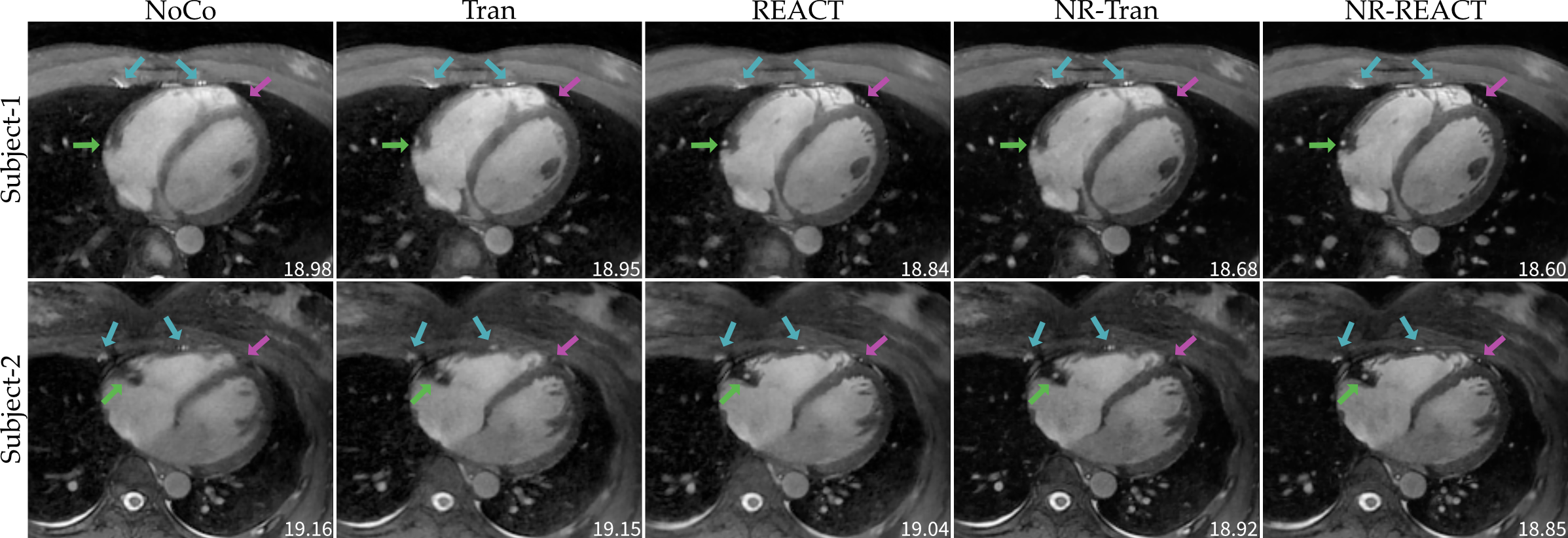}}
  \caption{Axial images comparing \texttt{NoCo}, \texttt{Tran}, \texttt{REACT}, \texttt{NR-Tran}, and \texttt{NR-REACT}; values denote image gradient entropy. Arrows highlight the left anterior descending artery (LAD, purple), right coronary artery (RCA, green), and internal thoracic vessels (blue). Compared with \texttt{Tran}, \texttt{REACT} substantially improved vessel depiction, whereas \texttt{NR-Tran} primarily enhanced overall sharpness with more limited gains in vessel delineation.}\label{fig:axials}
\end{figure*}

\begin{figure*}[tb]
  \centerline{\includegraphics[width=0.95\linewidth]{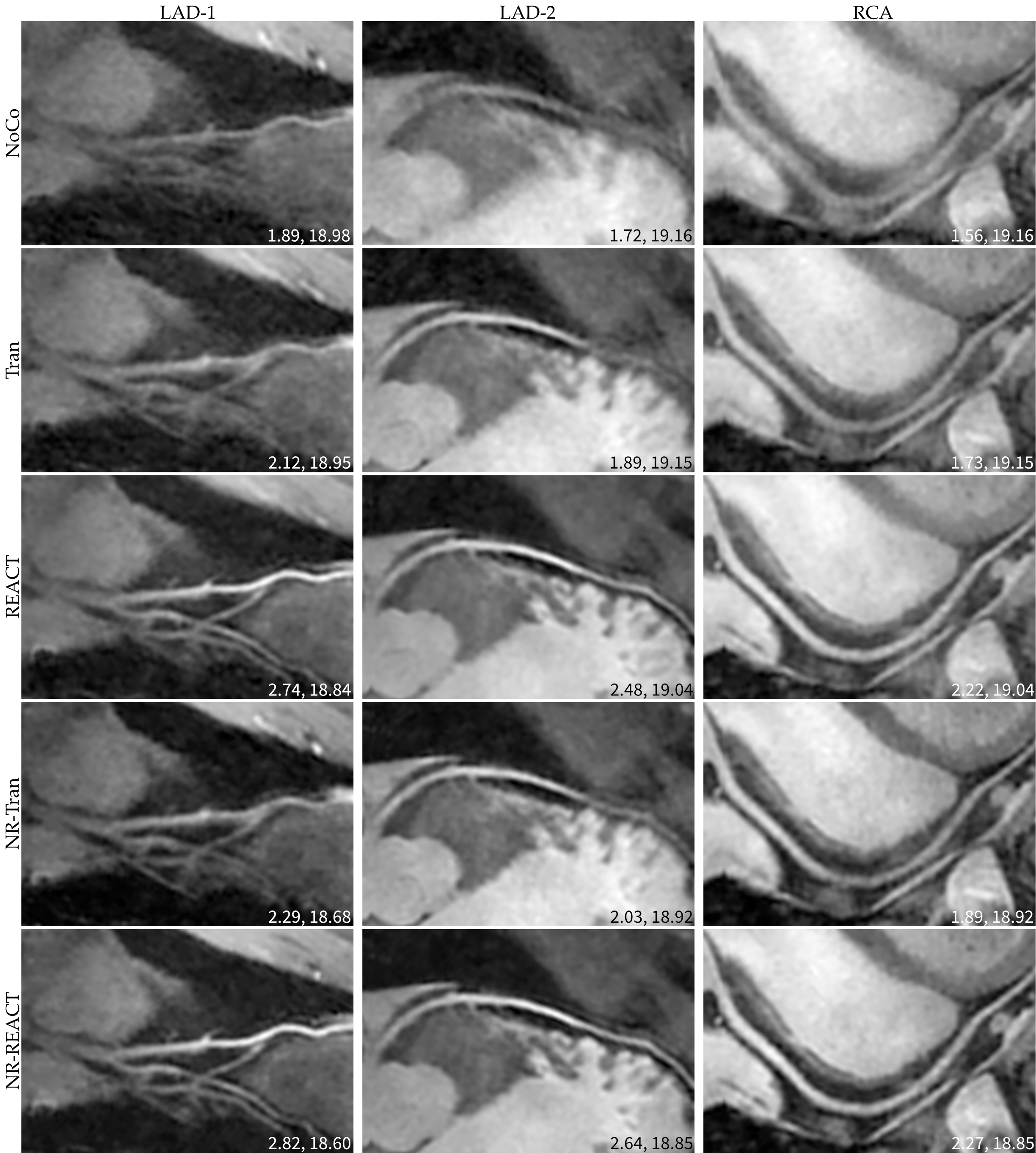}}
  \caption{Curved MPR images along the left anterior descending artery (LAD) and right coronary artery (RCA) comparing \texttt{NoCo}, \texttt{Tran}, \texttt{REACT}, \texttt{NR-Tran}, and \texttt{NR-REACT}; values denote unbounded image edge profile acutance (u-IEPA, left) and image gradient entropy (right). \texttt{REACT} showed improved LAD delineation compared to \texttt{Tran} and \texttt{NR-Tran}, while autofocus nonrigid refinement (\texttt{NR-Tran} and \texttt{NR-REACT}) yielded moderate additional sharpening over their respective baselines. Similar, less pronounced differences were observed for the RCA.}\label{fig:curved_mpr}
\end{figure*}

\begin{figure*}[tb]
  \centerline{\includegraphics[width=0.85\linewidth]{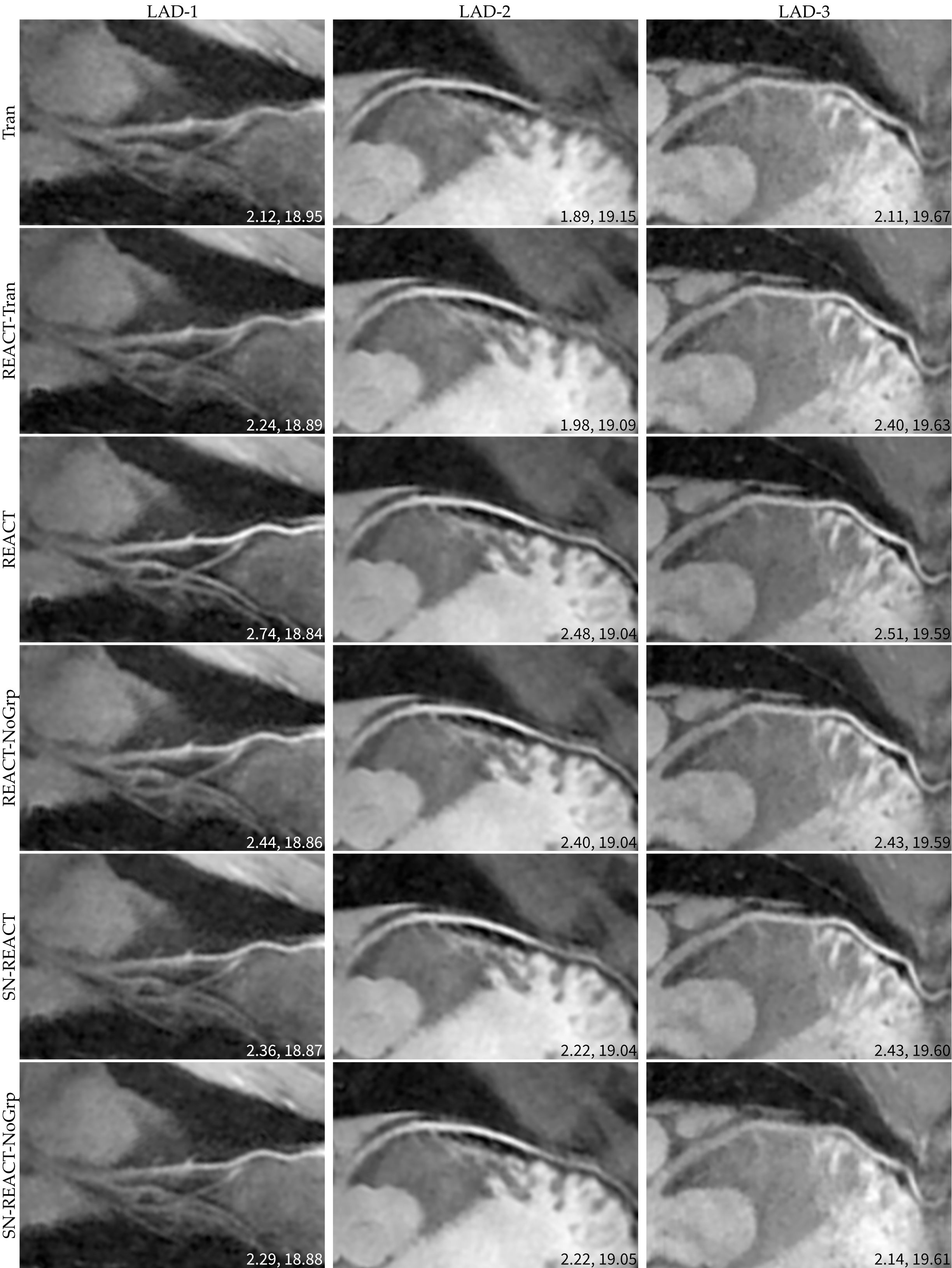}}
  \caption{Curved MPR images along the left anterior descending artery (LAD) comparing \texttt{Tran} and REACT variants; values denote unbounded image edge profile acutance (u-IEPA, left) and image gradient entropy (right). Improvements due to rotational motion correction (\texttt{REACT-Tran} vs.\ other REACT variants) were observed. \texttt{SN-REACT} outperformed \texttt{Tran} without using iNAVs. Incorporating group-wise updates in early iterations resulted in improved motion correction (\texttt{REACT-NoGrp} vs.\ \texttt{REACT} for LAD-1; \texttt{SN-REACT-NoGrp} vs.\ \texttt{SN-REACT} for LAD-3).}\label{fig:comp_inter}
\end{figure*}

Figure~\ref{fig:axials} compares axial images across the different motion-estimation methods. \texttt{REACT} substantially improved vessel depiction over \texttt{Tran}. Although \texttt{NR-Tran} enhanced overall image sharpness relative to \texttt{Tran}---including non-cardiac structures such as the internal thoracic vessels (blue arrows)---its improvement in vessel sharpness was more limited, especially for the LAD (purple arrows). Compared to \texttt{NR-Tran}, \texttt{REACT} yielded higher vessel acutance, whereas \texttt{NR-Tran} provided superior sharpness in the myocardium and non-cardiac regions. \texttt{NR-REACT} outperformed \texttt{NR-Tran} in both vessel visibility and overall image sharpness.

Since the motion correction used by \texttt{Tran} and \texttt{REACT} was performed in $k$-space, as described in \eqref{eq:traj_corr_ksp} and \eqref{eq:rigid_in_ksp}, the non-cardiac regions appeared blurrier than in \texttt{NoCo}. Because the autofocus nonrigid motion corrections were computed in a voxel-wise manner, the non-cardiac regions appeared notably sharper than in their respective baselines. However, \texttt{NR-Tran} provided more limited improvement in vessel depiction relative to \texttt{Tran} than \texttt{REACT} did, largely because its correction capability was constrained by the motion basis set. Rotational motion and finer translational motion were not effectively captured by the motion basis set, because it was constructed by scaling the estimated translational motion trajectory using a relatively sparse set of scale factors.

Figure~\ref{fig:curved_mpr} illustrates the curved MPR images along the LAD and RCA. For the LAD, \texttt{REACT} showed enhanced vessel delineation compared to \texttt{NR-Tran} and substantially improved depiction over \texttt{Tran}. Although the autofocus nonrigid motion corrections (\texttt{NR-Tran} and \texttt{NR-REACT}) yielded higher vessel sharpness than their respective baselines (\texttt{Tran} and \texttt{REACT}), these incremental gains were relatively moderate. For the RCA, the differences between the motion-estimation methods were less pronounced than for the LAD, though the overall trend remained consistent.

Figure~\ref{fig:comp_inter} compares curved MPR images along the LAD across the different REACT variants and \texttt{Tran}. The comparison between \texttt{REACT-Tran} and the other REACT methods indicates that correcting for rotational motion can considerably improve vessel acutance. \texttt{SN-REACT} outperformed \texttt{Tran} despite not utilizing iNAVs. Figure~S1 shows translation and rotation magnitudes estimated by \texttt{REACT} and \texttt{SN-REACT} for the three datasets corresponding to LAD~1--3.

\begin{figure*}[tb]
  \centerline{\includegraphics[width=0.95\linewidth]{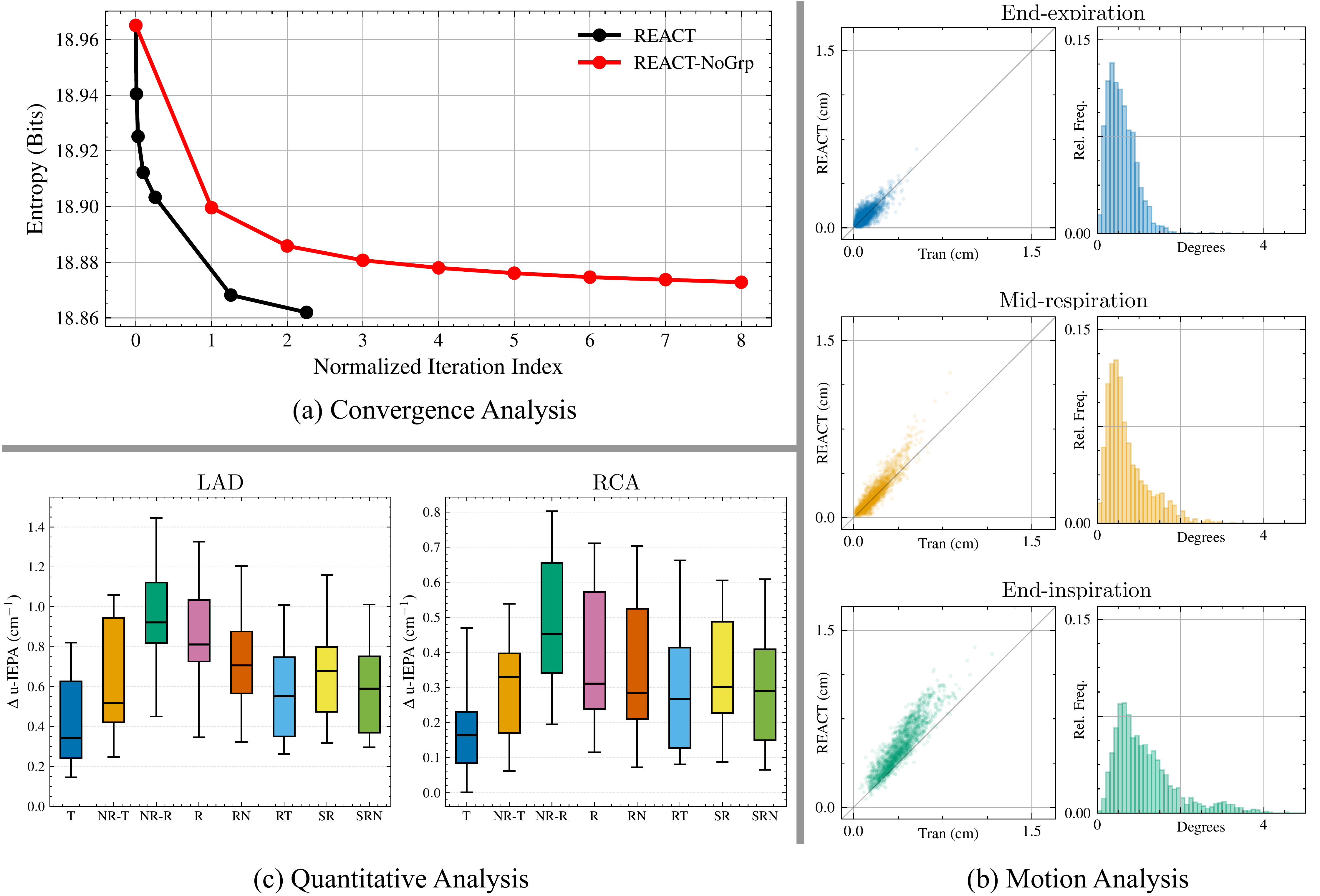}}
  \caption{(a) \textbf{Convergence analysis:} image gradient entropy versus normalized iteration index; \texttt{REACT} achieves lower entropy with fewer normalized iterations than \texttt{REACT-NoGrp}. (b) \textbf{Motion analysis:} translation magnitudes estimated by \texttt{Tran} (x-axis) and \texttt{REACT} (y-axis) across respiratory phases (end-expiration, mid-respiration, and end-inspiration),  with the identity line included for reference; rotation-angle distributions are also shown. In end-inspiration, most points lie above the diagonal, indicating larger translations estimated with \texttt{REACT}. (c) \textbf{Quantitative analysis:} boxplots of unbounded image edge profile acutance (u-IEPA) improvements relative to \texttt{NoCo} for the left anterior descending artery (LAD) and right coronary artery (RCA). Abbreviations: T = \texttt{Tran}, NR-T = \texttt{NR-Tran}, NR-R = \texttt{NR-REACT}, R = \texttt{REACT}, RN = \texttt{REACT-NoGrp}, RT = \texttt{REACT-Tran}, SR = \texttt{SN-REACT}, and SRN = \texttt{SN-REACT-NoGrp}.}\label{fig:three_analyses}
\end{figure*}

Figure~\ref{fig:three_analyses}(a) shows image-quality metrics plotted against normalized iteration indices for \texttt{REACT} and \texttt{REACT-NoGrp}.\footnote{The normalized interval for iteration $n$ was computed as $\sum_{m=1}^{M}|\mathbf{G}_{m}^{(n)}|/B$, the total number of subproblems across all partitions relative to a full coordinate-wise sweep; this quantity serves as a proxy for the computational load of REACT.} The LAD-1 dataset shown in Figure~\ref{fig:curved_mpr} was used for this analysis. Compared with \texttt{REACT-NoGrp}, \texttt{REACT} yielded lower image gradient entropy with fewer normalized iterations. The corresponding improvement in image quality for LAD-1 is shown in Figure~\ref{fig:comp_inter}. This improvement can be largely attributed to the correction of low-frequency trends in the early iterations through group-wise updates. Although \texttt{REACT} consistently achieved lower image gradient entropy at convergence than \texttt{REACT-NoGrp} across the other datasets, the corresponding improvements in vessel visibility were more modest for LAD-2 and LAD-3. A similar pattern was observed for \texttt{SN-REACT}: group-wise updates improved vessel depiction for LAD-3, whereas little qualitative difference was observed for LAD-1 and LAD-2.

Figure~\ref{fig:three_analyses}(b) shows scatter plots of translation magnitudes for three respiratory phases, with \texttt{Tran} on the x-axis and \texttt{REACT} on the y-axis. The respiratory phases were classified as end-expiration, mid-respiration, and end-inspiration using PCA features extracted from the iNAVs. In the end-inspiration phase, most points lie above the diagonal, indicating that \texttt{REACT} yielded larger translation estimates than \texttt{Tran}. The smaller translation estimates in \texttt{Tran} are likely due to mismatches between the end-expiration reference iNAV and the non-reference iNAVs. Because the heart can undergo rotations and nonrigid deformations across respiratory phases, iNAVs may not depict anatomy that differs only by translation; in our experiments, this mismatch appeared to result in conservative (i.e., smaller) translational shifts. Figure~\ref{fig:three_analyses}(b) also presents the distributions of rotation angles across respiratory phases. For most heartbeats, the estimated rotations were within $\pm 2^\circ$. Rotation angles tended to be slightly larger during the end-inspiration phase.

Figure~\ref{fig:three_analyses}(c) summarizes u-IEPA scores relative to \texttt{NoCo}; selected two-sided paired t-tests are reported in Table~\ref{tab:t-test}. 

\texttt{REACT} yielded higher u-IEPA for the LAD than \texttt{Tran} ($p<10^{-4}$). Although u-IEPA scores for the RCA were generally lower than those for the LAD and inter-method differences were smaller, \texttt{REACT} still provided higher u-IEPA than \texttt{Tran} ($p<0.005$). \texttt{SN-REACT} also showed higher u-IEPA than \texttt{Tran} for both the LAD ($p<0.001$) and RCA ($p<0.005$).

\texttt{REACT-Tran} resulted in higher u-IEPA than \texttt{Tran} for both the LAD ($p<0.005$) and RCA ($p<0.01$), indicating improved translational motion correction over the conventional iNAV-based method even without modeling rotation. \texttt{REACT} further achieved higher u-IEPA than \texttt{REACT-Tran} for both the LAD ($p<0.001$) and RCA ($p<0.05$), supporting the benefit of incorporating rotational motion.

\texttt{REACT} showed higher u-IEPA than \texttt{NR-Tran} for the LAD ($p<0.005$), consistent with the qualitative comparison. There was no significant difference in u-IEPA for the RCA ($p=0.10$). \texttt{NR-REACT} demonstrated higher u-IEPA than \texttt{NR-Tran} for both the LAD ($p<0.001$) and RCA ($p<0.005$).

Both \texttt{REACT} and \texttt{SN-REACT} yielded higher u-IEPA than their respective variants without group-wise updates for both the LAD and the RCA ($p<0.005$ for \texttt{REACT} and $p<0.05$ for \texttt{SN-REACT}).

\texttt{NR-Tran} and \texttt{NR-REACT} yielded considerably lower image gradient entropy than the REACT variants, with \texttt{NR-REACT} yielding lower entropy than \texttt{NR-Tran} ($p<10^{-4}$; Table~S2).


Although a relatively strict tolerance of 0.1~mm was used for the \verb|LN_BOBYQA| solver with search ranges of $\pm 2$~mm and $\pm 2^\circ$, \texttt{REACT} required an average of 27.37 cost function evaluations per subproblem, far fewer than an exhaustive grid search would require. Five iterations of \texttt{REACT} (equivalent to 1.29 normalized iterations) required 1.21~hours, whereas two iterations of \texttt{REACT-NoGrp} required 1.58~hours. The total elapsed time for \texttt{SN-REACT} was 5.47~hours.

\begin{table*}[t]%
  \caption{Two-sided Paired t-tests for Selected Comparisons.\label{tab:t-test}}
  \vspace{6pt}
  {%
  \begin{tabular*}{\textwidth}{@{\extracolsep{\fill}} ll cc cc @{}}
    \toprule
    \multicolumn{2}{c}{\textbf{Comparisons} (A $\neq$ B)} 
    & \multicolumn{2}{c}{\textbf{u-IEPA (LAD)}} 
    & \multicolumn{2}{c}{\textbf{u-IEPA (RCA)}}\\
    \cmidrule(lr){1-2} \cmidrule(lr){3-4} \cmidrule(lr){5-6}
    \textbf{A}      & \textbf{B} 
    & \textbf{Diff} & \textbf{p-value} 
    & \textbf{Diff} & \textbf{p-value} \\
    \midrule
    Tran       & NoCo           & 0.425 $\pm$ 0.251 & 4.6e-4 & 0.175 $\pm$ 0.140 & 3.3e-3 \\
    NR-Tran    & Tran           & 0.201 $\pm$ 0.077 & 1.8e-5 & 0.127 $\pm$ 0.077 & 5.5e-4 \\
    REACT      & Tran           & 0.405 $\pm$ 0.154 & 1.7e-5 & 0.205 $\pm$ 0.144 & 1.5e-3 \\
    SN-REACT   & Tran           & 0.266 $\pm$ 0.146 & 2.8e-4 & 0.161 $\pm$ 0.128 & 3.2e-3 \\
    REACT-Tran & Tran           & 0.141 $\pm$ 0.099 & 1.5e-3 & 0.128 $\pm$ 0.113 & 6.1e-3 \\
    REACT      & REACT-Tran     & 0.264 $\pm$ 0.165 & 6.8e-4 & 0.077 $\pm$ 0.084 & 1.8e-2 \\
    REACT      & NR-Tran        & 0.203 $\pm$ 0.168 & 4.1e-3 & 0.078 $\pm$ 0.136 & 1.0e-1 \\
    NR-REACT   & NR-Tran        & 0.318 $\pm$ 0.174 & 2.7e-4 & 0.181 $\pm$ 0.126 & 1.4e-3 \\
    REACT      & REACT-NoGrp    & 0.102 $\pm$ 0.084 & 4.0e-3 & 0.038 $\pm$ 0.030 & 2.8e-3 \\
    SN-REACT   & SN-REACT-NoGrp & 0.088 $\pm$ 0.089 & 1.2e-2 & 0.037 $\pm$ 0.050 & 4.2e-2 \\
    \bottomrule
  \end{tabular*}
  }%
  \begin{tablenotes}
    \item Values represent the mean $\pm$ standard deviation of metric differences (A - B). 
  \end{tablenotes}
\end{table*}

\section{Discussion}\label{sec:discussion}

We have presented a new autofocus method, REACT, for estimating 3D rigid motion in MR imaging. REACT employs coordinate descent to decompose the high-dimensional optimization problem into a series of subproblems, each updating the motion parameters for a single temporal segment while keeping the others fixed. Assuming the cost functions of these subproblems are approximately locally convex, each subproblem is efficiently solved using a derivative-free solver. We incorporate group-wise updates in the early iterations to address low-frequency errors in the initial motion estimates. When accurate initial motion estimates are unavailable, REACT uses successive 1D sweeps and a learning-rate ramp-up to improve robustness.

Our approach is computationally feasible, largely due to recent advances in derivative-free optimization methods such as the BOBYQA algorithm~\cite{powell2009bobyqa,johnson_nlopt}. In the \textit{in vivo} experiments, only a few dozen cost function evaluations were required to solve each coordinate-descent subproblem, which is substantially fewer than would be required for an exhaustive grid search. Together with the numerical simulations, this finding is consistent with our hypothesis that the objective functions of the subproblems are approximately locally convex. Because of this local convexity, each subproblem can be solved efficiently using quadratic approximations of the cost function within the BOBYQA solver.

In principle, the proposed approach could incorporate analytic derivatives of the image-quality metric with respect to the motion parameters, as in the method of Loktyushin et al.~\cite{loktyushin_blind_2013}. Although a gradient-based solver may reduce the number of cost function evaluations per subproblem, deriving and computing these derivatives can be nontrivial and computationally expensive for non-Cartesian acquisitions, where gridding usually relies on convolutional interpolation~\cite{bernstein2004handbook, beatty_rapid_2005}. Therefore, this study prioritized a derivative-free approach to ensure broader applicability and ease of implementation. Future work could evaluate the relative performance and computational efficiency of different numerical solvers for REACT, including derivative-free and gradient-based solvers.

The correction of rotational motion is often neglected in free-breathing cardiac MRI reconstructions. This may be partly because, although rotational motion is more difficult to estimate from limited navigator data, translational motion correction often yields sufficiently good results. In our experiments, translational motion dominated the overall respiratory motion; however, incorporating rotational motion also provided noticeable improvements in image quality and vessel depiction. It may be beneficial to incorporate autofocus motion correction into the reconstruction pipeline to address rotational motion when additional computation time is available.

In this work, no regularization was applied for motion estimation because the motion between neighboring heartbeats often differs due to the relatively large temporal spacing between them. We did not observe any noticeable outlier motion estimates even without regularization. However, when the motion between neighboring temporal segments is smooth, it may be beneficial to include an additional term that penalizes large deviations among neighboring segments. In fact, the coordinate-descent methods for tomographic image reconstruction readily incorporated generalized Gaussian Markov random field (GGMRF) models for regularization~\cite{bouman_unified_1996, ye_optical_1999, yu_fast_2011}, thanks to its pixel-wise update scheme.



To reduce the computational cost of objective function evaluations, we used the SoS reconstruction. In our implementation, only a small portion of the data was gridded on the fly while the static background was precomputed, which significantly accelerated the SoS reconstruction. However, the SoS method may suffer from aliasing artifacts when the data are highly undersampled. Future work could explore integrating more advanced reconstruction methods, such as deep learning-based techniques~\cite{heckel2024deep}.



While \texttt{SN-REACT} demonstrated promising results without requiring initial motion estimates, the proposed approach can still benefit from initialization or other external information. Future work could explore motion-estimation methods that provide sufficiently accurate initial estimates while having minimal impact on data acquisition.

\section{Conclusion}\label{sec:conclusion}

This study demonstrates the feasibility of the coordinate-descent approach for autofocus motion correction in MR imaging. The high-dimensional optimization problem is decomposed into a series of subproblems, each updating the motion parameters of a single temporal segment while keeping the others fixed. Our observations indicate that, provided the data acquisition satisfies certain conditions and the current motion estimate is sufficiently close to the desired solution, the cost functions of these subproblems exhibit approximate local convexity. Consequently, each coordinate-descent subproblem is solved efficiently using a derivative-free numerical solver, avoiding an exhaustive grid search. Experiments in free-breathing coronary MR angiography demonstrate that REACT accurately estimates rigid respiratory motion, yielding improved image quality and enhanced coronary artery depiction.

\section*{Acknowledgments}

The authors thank Dr.~Mario~O.~Malav\'{e} for providing the \textit{in vivo} datasets used in this study. This work was supported in part by NIH grant R01~HL127039.

\bibliographystyle{ieeetr}
\bibliography{references}

\clearpage

\newcommand{\supplementtitle}{%
  \begin{center}
    \LARGE\bfseries
    Supporting Information for Rigid Motion Estimation using
    Accelerated Coordinate Descent (REACT) for MR Imaging\par
  \end{center}
  \vspace{1em}
}
\makeatletter
\if@twocolumn
  \twocolumn[\supplementtitle]
\else
  \supplementtitle
\fi
\makeatother

\setcounter{section}{0}
\setcounter{figure}{0}
\setcounter{table}{0}

\renewcommand{\thesection}{S\arabic{section}}
\renewcommand{\thesubsection}{\thesection.\arabic{subsection}}

\renewcommand{\thesectiondis}{\thesection}
\renewcommand{\thesubsectiondis}{\thesubsection}

\renewcommand{\thefigure}{S\arabic{figure}}
\renewcommand{\thetable}{S\arabic{table}}
\captionsetup[figure]{name=Figure}

\begin{bibunit}[ieeetr]
\section{Setups of Numerical Simulations}\label{supp:sec:method-simul}

The common simulation setup---sampling trajectories, noise, and coil sensitivities---and the procedures for simulating motion are described first. The specific parameters and implementation details for each study presented in Figures 2--4 are then provided.


\subsection{Sampling Trajectories}\label{supp:sec:method-simul:traj}

Two 2D acquisition schemes were simulated: interleaved multi-shot echo-planar imaging (EPI) and interleaved spiral imaging. For the spiral trajectory, parameters similar to those described by Kasper et al.~\cite{kasper_rapid_2018} were used: a field of view (FOV) of 23~cm, a resolution of 0.5~mm, 30 interleaves, a readout duration of 20~ms, a total scan duration of 90~s, a maximum gradient amplitude of 31~mT/m, and a maximum slew rate of 200~mT/m/ms. The interleaved EPI trajectories were constructed using a simple procedure in which evenly spaced $k_y$-lines were selected for each shot.

\subsection{Noise and Coil Sensitivity}\label{supp:sec:method-simul:noise}

The signal-to-noise ratio (SNR) was selected based on the measurements reported by de Zwart et al.~\cite{zwart_snr_2004}. When their 16-channel brain coil was used on a 3.0~T scanner with a voxel size of $1.7$~mm $\times$ $1.7$~mm $\times$ $3.5$~mm, the mean SNR over the whole brain was about 120. In our simulations, a voxel size of $0.5$~mm $\times$ $0.5$~mm $\times$ $2.7$~mm was assumed, and the SNR was set to 8. Complex Gaussian noise was added to the $k$-space data to set the SNR of the sum-of-squares (SoS) reconstruction to 8, measured within the elliptical region of interest (ROI) enclosing the brain. A 16-channel receive array with linear sensitivities was simulated.

\subsection{Uncorrelated Motion Model}\label{supp:sec:method-simul:uncorrelated}

Each component of 2D rigid motion was sampled independently from a uniform distribution. Translational motion was simulated as follows:
\begin{align}
    \mathbf{x}_b &\sim \mathcal{U}(-u_t, u_t),\\
    \mathbf{y}_b &\sim \mathcal{U}(-u_t, u_t),
\end{align}
\noindent where $u_t$ represents the upper bound on the magnitude of each translational component, and $\mathbf{x}_b$ and $\mathbf{y}_b$ denote the translations along the $x$- and $y$-directions, respectively, at the $b$-th beat.

To express translational and rotational motion on a common spatial scale, the amount of rotational motion was represented by the corresponding arc length at the boundary of a disk of radius $r = 6.5$~cm. The upper bound on this arc length was set to half the maximum translational amplitude to simulate predominantly translational motion. The rotation angles were therefore generated as follows:
\begin{align}
    u_\theta &= \frac{u_t}{2 r},\\
    \bm{\theta}_b &\sim \mathcal{U}(-u_\theta, u_\theta),
\end{align}
\noindent where $\bm{\theta}_b$ denotes the rotation angle at the $b$-th beat, expressed in radians.

\subsection{Correlated Motion Model}\label{supp:sec:method-simul:correlated}

The root-mean-square (RMS) deviation for 3D rigid motion~\cite{jenkinson1999rms} was adapted to quantify the magnitude of 2D rigid motion between two adjacent beats as follows:
\begin{align}
    \mathbf{A} &= \mathbf{R}_{\text{2D}}(\Delta\bm{\theta}_b) - \mathbf{I}_{\text{2D}},\\
    \text{RMS}_{\text{2D}}(\Delta\mathbf{t}_b, \Delta\bm{\theta}_b) &= \sqrt{\frac{r^2}{4}\text{tr}\left(\mathbf{A}^T\mathbf{A}\right) + \Delta\mathbf{t}_b^T \Delta\mathbf{t}_b}\\
    &= \sqrt{r^2(1 - \cos\Delta\bm{\theta}_b) + \lVert \Delta\mathbf{t}_b \rVert^2},\label{eq:rms_dev_2d}
\end{align}
\noindent where $\Delta\mathbf{t}_b$ and $\Delta\bm{\theta}_b$ denote the changes in translation and rotation angle, respectively, from the $(b\!-\!1)$-th beat to the $b$-th beat; $\mathbf{I}_{\text{2D}}$ and $\mathbf{R}_{\text{2D}}(\Delta\bm{\theta}_b)$ represent the identity and rotation matrices, respectively; and $r$ denotes the radius of the disk.

A Gaussian random walk model was employed to simulate slowly drifting motion, as often observed in brain MRI~\cite{maclaren_measurement_2012,schilling_head_2025}. The unscaled change in each motion component at the $b$-th beat was sampled independently from a normal distribution:
\begin{align}
    \Delta\tilde{\mathbf{x}}_b &\sim \mathcal{N}(0, 1),\\
    \Delta\tilde{\mathbf{y}}_b &\sim \mathcal{N}(0, 1),\\
    \Delta\tilde{\bm{\theta}}_b &\sim \mathcal{N}(0, 1/(2r^2)).
\end{align}
\noindent To simulate predominantly translational motion, the rotational variance was scaled to yield approximately half the RMS displacement of either translational component under the small-angle approximation.

The unscaled RMS displacement per minute (denoted RMSpm in~\cite{reuter_head_2015}) was then computed using \eqref{eq:rms_dev_2d} as follows:
\begin{align}
    \tilde{D} &= \frac{1}{T}\sum_{b=1}^B \text{RMS}_{\text{2D}}(\Delta\tilde{\mathbf{t}}_b, \Delta\tilde{\bm{\theta}}_b)\\
    &=\frac{1}{T}\sum_{b=1}^B \sqrt{r^2(1 - \cos \Delta\tilde{\bm{\theta}}_b) + \lVert \Delta \tilde{\mathbf{t}}_b \rVert^2},
\end{align}
\noindent where $T$ denotes the total scan time in minutes.

The translation along the $x$-direction was computed as the cumulative sum of the unscaled changes and subsequently scaled using the target RMSpm, $D$, as follows:
\begin{align}
    \tilde{\mathbf{x}}_b &= \sum_{i=1}^b \Delta\tilde{\mathbf{x}}_i,\\
    \mathbf{x}_b &= \left(\frac{D}{\tilde{D}}\right)\tilde{\mathbf{x}}_b.
\end{align}
\noindent $\mathbf{y}_b$ and $\bm{\theta}_b$ were computed using a similar procedure.

\subsection{Figure 2: Objective Surfaces of Coordinate-Wise Cost Functions}\label{supp:sec:method-simul:fig2}

The uncorrelated motion model described in~\ref{supp:sec:method-simul:uncorrelated} was used to simulate 2D translational motion. Rotational motion was omitted from this simulation so that the objective surfaces could be visualized as 2D maps. The maximum translation amplitude ($u_t$) was arbitrarily set to 2~mm. The number of $k_y$-lines per shot was set to 12 for the EPI acquisition. The spiral trajectory described in~\ref{supp:sec:method-simul:traj} was used for the spiral acquisition. At each iteration, the cost function was sampled on a $64 \times 64$ grid spanning $\pm 5$~mm in each direction around the ground truth, and the resulting cost maps were averaged over beats. Because the image gradient entropy in~(3) is insensitive to global translation, the motion-corrected image may be shifted relative to the ground truth. To generate centered cost maps, the translation estimates were shifted to have zero mean before the cost maps were computed.

The initial 2D translation estimates were set to zero, and three REACT iterations were performed. A linear learning-rate ramp-up was used. Initial 1D sweeps were performed over a range of $\pm 1.25 u_t$ with a step size of 0.5~mm (i.e., the pixel size), starting with the $x$-direction. The tolerance and search range for the numerical solver were set to $0.001$~mm and $\pm 2$~mm, respectively.

\subsection{Figure 3: Comparing REACT across Sampling Trajectories and Motion Amplitudes}\label{supp:sec:method-simul:fig3}

The uncorrelated motion model described in~\ref{supp:sec:method-simul:uncorrelated} was used to simulate 2D rigid motion. A wide range of motion amplitudes was tested, with $u_t$ varied from 1.0~mm to 4.0~mm at 0.5~mm intervals.

Four sampling trajectories were tested: two interleaved EPI and two interleaved spiral trajectories. For the EPI acquisitions, 8 and 16 $k_y$-lines per shot were used; twice as many PE lines were collected per shot in the latter as in the former. The two spiral trajectories were modified versions of the trajectory described in~\ref{supp:sec:method-simul:traj}. For the first trajectory, the readout duration was halved to 10~ms, and the number of interleaves was doubled accordingly, resulting in sparser $k$-space coverage per interleaf. For the second trajectory, the original 20~ms readout duration was retained, and the variable-density sampling parameter~\cite{kim2003simple} was set to $\alpha=1.5$ to increase the sampling density near the center of $k$-space.


For each trajectory and each motion magnitude, 25 simulations were conducted; for each run, a different motion trajectory and different $k$-space noise were generated. The cases whose structural similarity (SSIM) indices were closest to the corresponding mean SSIM were selected as representative cases for each $u_t$.

The initial motion estimates were set to zero, and 10 REACT iterations were performed. The learning rate was increased linearly to 1.0 over the first 7 iterations. Initial 1D sweeps were performed over a range of $\pm 1.25 u_t$ with a step size of 0.5~mm (i.e., the pixel size), starting with the $x$-direction. The tolerance of the numerical solver was set to $0.001$~mm. The search ranges of the numerical solver were set to $\pm u_t/4$ for translation and $\pm u_\theta$ for rotation.

\subsection{Figure 4: Group-Wise Updates in the Presence of Low-Frequency Errors}\label{supp:sec:method-simul:fig4}

The correlated motion model described in~\ref{supp:sec:method-simul:correlated} was used to simulate 2D rigid motion. The RMS displacement per minute was set to 9~mm/min.

The spiral trajectory with a 10~ms readout duration, described in~\ref{supp:sec:method-simul:fig3}, was used. The total scan duration was set to 180~s ($2 \times 90$~s) to reflect the doubled number of interleaves.

For group-wise updates, the temporal segments were initially divided into three contiguous groups. The number of groups was then doubled at each iteration, approximately halving the group size, over five iterations. At each iteration, three partitions ($M\!=\!3$) with different starting offsets were used to soften group boundaries.

Parameters similar to those described in~\ref{supp:sec:method-simul:fig3} were used. The translation and rotation ranges ($u_t$ and $u_\theta$) were estimated from 25 realizations of the correlated motion.
\section{Setups of the \textit{In Vivo} Study}\label{supp:sec:method-invivo}

\subsection{Partitioning}\label{supp:sec:method-invivo:partition}

For the group-wise update in REACT, the temporal segments were classified into multiple respiratory phases using recursive k-means clustering. The number of clusters was initially set to 3 and progressively increased over four iterations. At each iteration, the number of clusters was determined numerically so that the new largest cluster was approximately half the size of the largest cluster in the previous iteration. To reduce sensitivity to group boundaries, three partitions ($M\!=\!3$) were used: one with the numerically determined number of clusters and the other two with one fewer and one more cluster, respectively.

During the initial clustering, the cluster with the smallest standard deviation about its centroid was selected as the reference. In subsequent clustering steps, the reference cluster was selected based on the proximity of its centroid to the previous reference centroid. The clusters were ordered by the distances of their centroids from the reference centroid, and the group-wise updates were performed from the farthest cluster to the nearest. The reference cluster was held fixed during these updates to serve as the static background. In the fifth and sixth iterations, coordinate-wise updates were performed in a randomized order.

\subsection{Parameters of REACT and SN-REACT}

For \texttt{REACT}, the number of iterations was set to 5; group-wise updates were used during the first 4 iterations, as described in~\ref{supp:sec:method-invivo:partition}. The numerical solver tolerance was set to $0.1$~mm, and the search ranges for translation and rotation were set to $\pm 2$~mm and $\pm 2^\circ$, respectively. 1D sweeps and learning-rate ramp-up were not used.

For \texttt{SN-REACT}, the number of outer iterations was set to 5. During the first 3 iterations, only 3D translational motion was estimated, and both a linear learning-rate ramp-up and 1D sweeps were applied. The 1D sweeps were performed along the superior--inferior (SI), anterior--posterior (AP), and left--right (LR) directions, in that order, with respective ranges of $\pm 10$, $\pm 4$, and $\pm 4$~mm and a step size of $2$~mm. During the last 2 iterations, 3D rigid motion was estimated. To reduce computation time, the 1D sweep was performed only along a principal direction over a range of $\pm 5$~mm at $1$~mm intervals. The number of inner iterations was set to 3: the first 2 for group-wise updates and the last 1 for coordinate-wise updates. The numerical solver tolerance was set to $0.1$~mm, and the search ranges for translation and rotation were set to $\pm 2$~mm and $\pm 2^\circ$, respectively.

For \texttt{SN-REACT-NoGrp}, parameters similar to those for \texttt{SN-REACT} were used, but without inner iterations. The number of iterations was increased to 6, with the first 3 retained for translational motion estimation and 1 additional iteration used for rigid motion estimation. The additional iteration was included to make the computation time comparable to that of \texttt{SN-REACT}.

\subsection{Initial Translational Motion Estimation}

Initial estimates of translational motion (\texttt{Tran}) were obtained by registering each image-based navigator (iNAV) to a reference frame. For datasets with 3D iNAVs, 3D translational registration implemented in SimpleITK~\cite{lowekamp2013design, mccormick2014itk} was applied. For datasets with 2D iNAVs, 2D affine registration was employed instead, since the imaged slice of the heart could vary between heartbeats. A cross-correlation-based subpixel registration method~\cite{guizar-sicairos_efficient_2008} was used to provide an initial guess for the SimpleITK registrations.

\subsection{Conventional Autofocus Nonrigid Motion Correction}

For the autofocus nonrigid motion correction methods (\texttt{NR-Tran} and \texttt{NR-REACT}), parameters similar to those used by Ingle et al.~\cite{ingle_nonrigid_2014} were employed. Evenly spaced scale factors from $0\times$ to $2\times$ were used to scale the estimated translational motion trajectories: 9 along SI, 9 along the axis (AP or LR) with the larger motion amplitude, and 5 along the remaining axis~\cite{ingle_nonrigid_2014}. An isotropic window of 32 voxels (3.84~cm) weighted by a Hann window~\cite{cheng_nonrigid_2012,ingle_nonrigid_2014} was used to evaluate the local image gradient entropy. TV denoising was applied to the assembled SoS reconstruction using the Split Bregman algorithm~\cite{goldstein_split_2009}. For \texttt{NR-REACT}, rotational motion was not incorporated into the motion basis set; instead, rotational motion correction was applied to each basis image according to (1).

\subsection{Final Motion-Corrected Reconstruction}

For the final image reconstruction, the Split Bregman algorithm~\cite{goldstein_split_2009} with total variation (TV) regularization was applied, using coil sensitivity maps estimated by ESPIRiT~\cite{uecker_espiriteigenvalue_2014}.

\section{Additional Results}\label{supp:sec:results}

\begin{figure*}[p]
  \centerline{\includegraphics[width=0.75\linewidth]{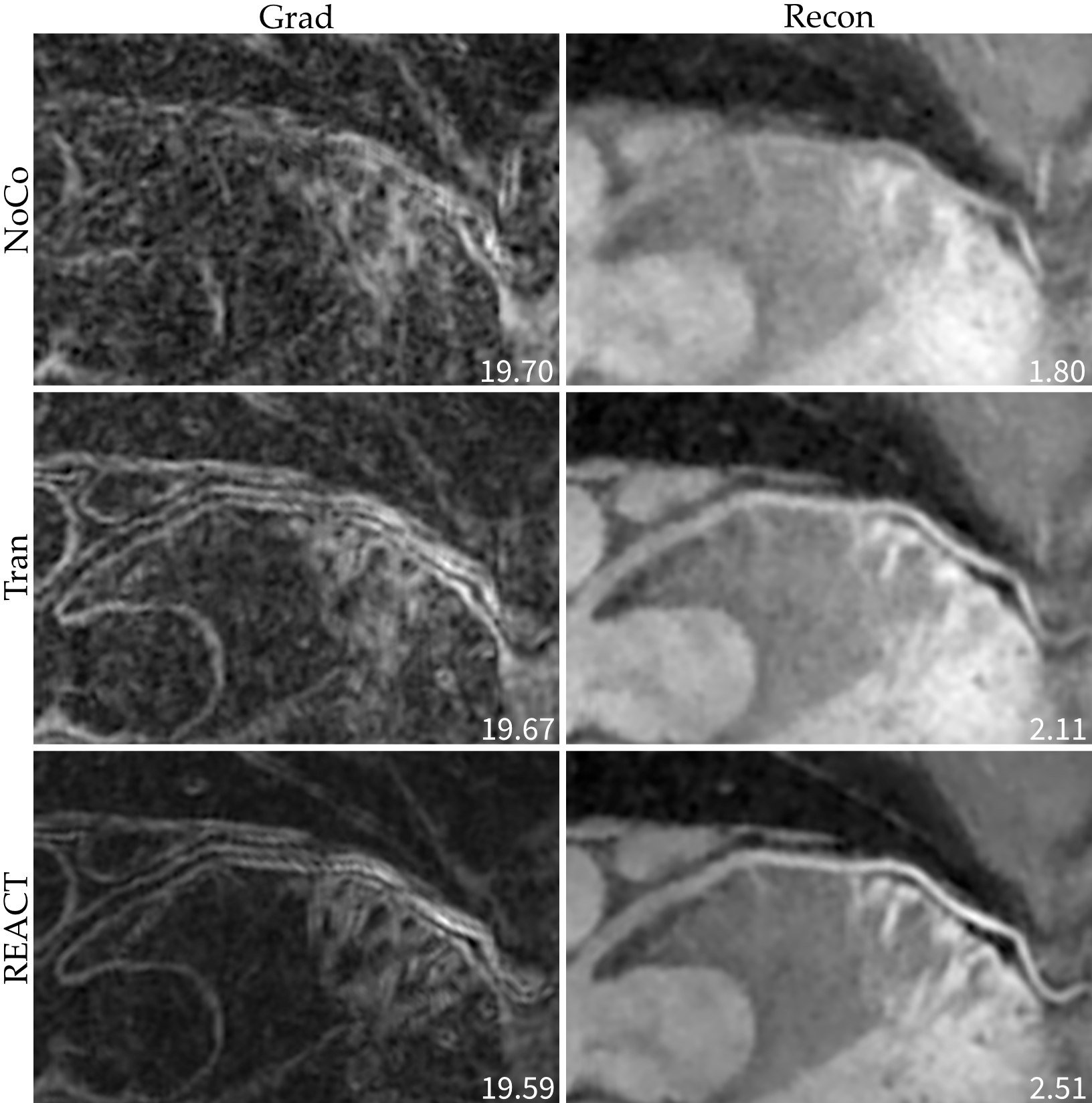}}
  \caption{Curved MPR images along the left anterior descending artery (LAD) for \texttt{NoCo} (top), \texttt{Tran} (middle), and \texttt{REACT} (bottom). Left: image-gradient magnitudes ($\mathbf{H}$ in (5)); values denote image gradient entropy. Right: corresponding reconstructions; values denote unbounded image edge profile acutance (u-IEPA). Compared with \texttt{NoCo} and \texttt{Tran}, \texttt{REACT} suppressed background graininess and produced sharper, more continuous vessel edges in the image-gradient magnitude map, resulting in improved vessel visibility in the reconstruction.}\label{supp:fig:entropy}
\end{figure*}

Figure~\ref{supp:fig:entropy} shows curved multi-planar reformation (MPR) images along the left anterior descending artery (LAD) and the corresponding image gradient magnitudes defined in (5). Compared with \texttt{NoCo} and \texttt{Tran}, \texttt{REACT} suppressed high-frequency background graininess and produced sharper, more continuous vessel edges in the image gradient magnitude map. These changes in the image gradient field translated into improved vessel sharpness and visibility in the motion-corrected reconstruction.

\begin{figure*}[p]
  \centerline{\includegraphics[width=0.975\linewidth]{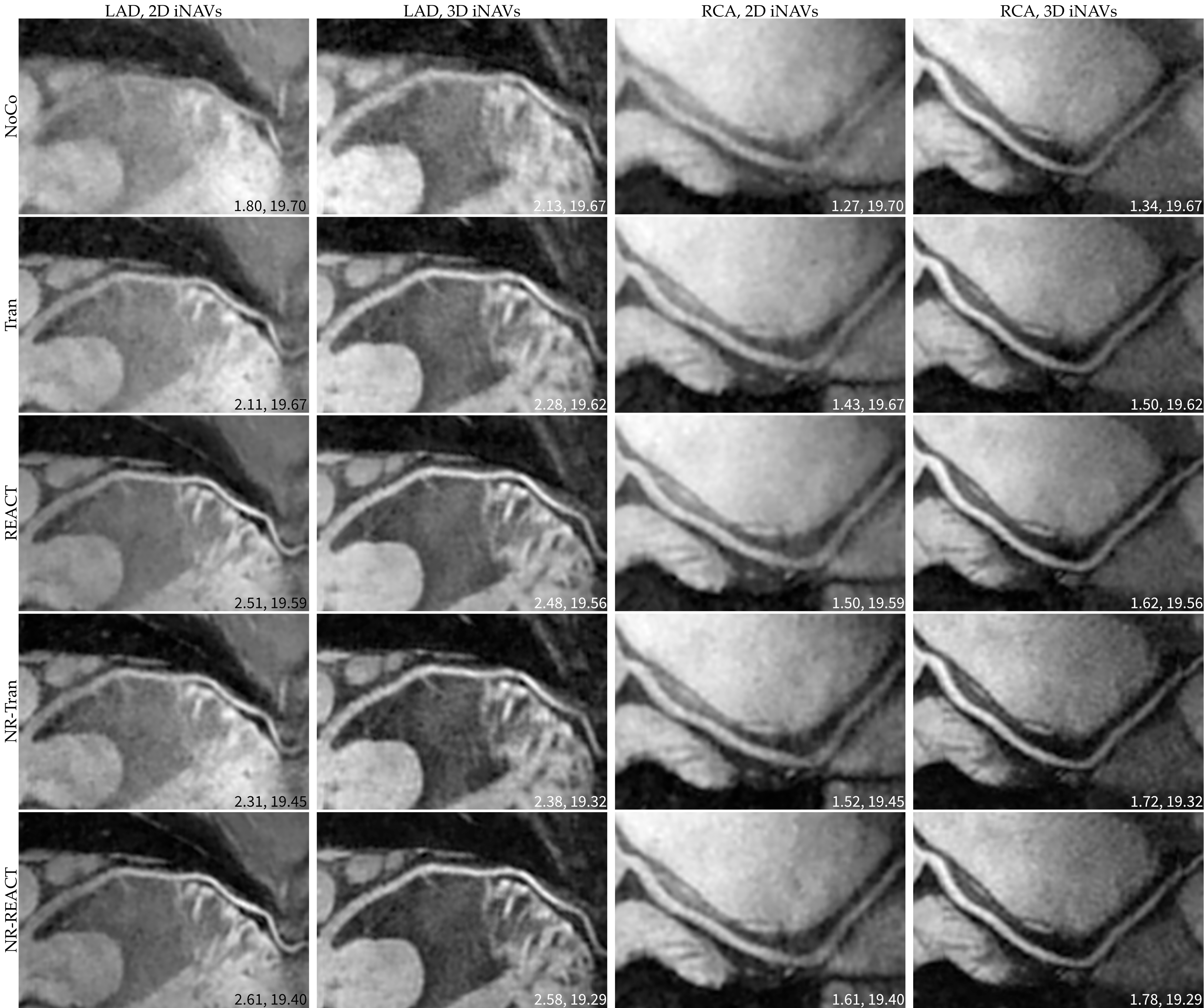}}
  \caption{Curved MPR images along the left anterior descending artery (LAD) and right coronary artery (RCA) comparing \texttt{NoCo}, \texttt{Tran}, \texttt{REACT}, \texttt{NR-Tran}, and \texttt{NR-REACT}; values denote unbounded image edge profile acutance (u-IEPA, left) and image gradient entropy (right). Two datasets were acquired from the same subject using ATR-bSSFP sequences: one with leading and trailing 2D iNAVs~\cite{wu2013free} and the other with trailing 3D iNAVs~\cite{addy20173d}. Both datasets exhibited similar trends: \texttt{REACT} improved LAD delineation compared with \texttt{Tran} and \texttt{NR-Tran}, while autofocus nonrigid refinement (\texttt{NR-Tran} and \texttt{NR-REACT}) provided moderate additional sharpening over the respective baselines. Differences among methods were less pronounced for the RCA; in this subject, \texttt{NR-Tran} yielded slightly higher RCA u-IEPA than \texttt{REACT} with both navigator types.}
  \label{supp:fig:mpr_inavs}
\end{figure*}

The datasets shown in Figures 5--7 were acquired with 2D iNAVs. Figure~\ref{supp:fig:mpr_inavs} shows curved MPR images from two datasets acquired from the same subject using 2D and 3D iNAVs. Both datasets showed similar trends across the motion-correction methods, with greater improvements in LAD delineation and more modest improvements for the right coronary artery (RCA).

\begin{figure*}[p]
  \centerline{\includegraphics[width=0.9\linewidth]{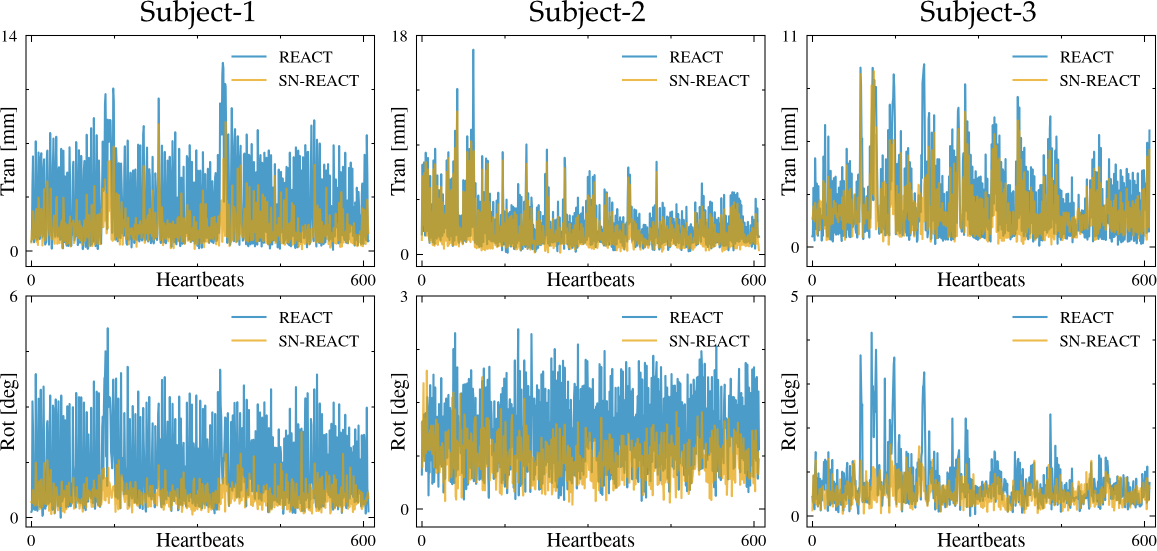}}
  \caption{Translation and rotation magnitudes estimated by \texttt{REACT} (blue) and \texttt{SN-REACT} (orange). The norms of the translation vectors ($\lVert \mathbf{t} \rVert$, top row) and rotation vectors ($\lVert \mathbf{v} \rVert$, bottom row) are plotted across heartbeats. Columns correspond to the three datasets shown as LAD~1--3 in Figure~7, from left to right. \texttt{SN-REACT} showed closer agreement with \texttt{REACT} for Subjects~2 and~3 than for Subject~1, which exhibited the largest mean translation magnitude and rotation angle. For Subjects~1--3, the mean translation magnitudes were 3.08, 2.40, and 2.28~mm, and the mean rotation angles were $1.32^\circ$, $1.10^\circ$, and $0.75^\circ$, respectively. The corresponding LAD u-IEPA scores were 2.74, 2.48, and 2.51 for \texttt{REACT} and 2.36, 2.22, and 2.43 for \texttt{SN-REACT}, respectively.}
  \label{supp:fig:sn_vs_react_mag}
\end{figure*}

Figure~\ref{supp:fig:sn_vs_react_mag} compares translation and rotation magnitudes estimated by \texttt{REACT} and \texttt{SN-REACT}. \texttt{SN-REACT} showed closer agreement with \texttt{REACT} for subjects with smaller mean translation and rotation magnitudes.

\begin{figure*}[p]
  \centerline{\includegraphics[width=0.6\linewidth]{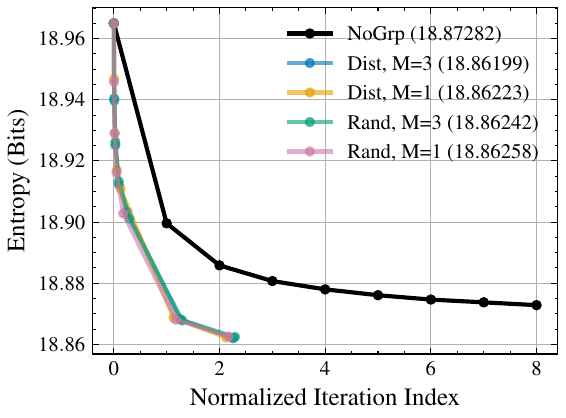}}
  \caption{Image gradient entropy versus normalized iteration index for different update ordering schemes (farthest-first or randomized) and different numbers of partitions. Dist and Rand denote farthest-first and randomized ordering for group-wise updates, respectively. $M$ denotes the number of partitions used at each iteration. NoGrp denotes \texttt{REACT-NoGrp}. Values in parentheses indicate image gradient entropy at the final iteration. Although the settings used in the \textit{in vivo} study (Dist, $M\!=\!3$) yielded slightly lower final image gradient entropy than the others, similar convergence behavior was observed across the four configurations.}\label{supp:fig:partition}
\end{figure*}

Figure~\ref{supp:fig:partition} compares convergence behavior with different update ordering schemes (farthest-first or randomized) and different numbers of partitions. Although the settings used in the \textit{in vivo} study (farthest-first, $M\!=\!3$) yielded slightly lower final image gradient entropy than the others, similar convergence behavior was observed across the four configurations.

The results of selected two-sided paired t-tests for image gradient entropy reduction are reported in Table~\ref{supp:tab:t-test-entropy}. u-IEPA quantifies coronary artery sharpness, whereas image gradient entropy serves as a measure of overall image sharpness within the ROI.

The autofocus nonrigid motion correction methods (\texttt{NR-Tran} and \texttt{NR-REACT}) yielded considerably lower image gradient entropy than the REACT variants. This may reflect differences in their correction strategies: REACT estimates motion by minimizing image gradient entropy across the ROI as a whole, whereas Ingle et al.'s method uses localized image gradient entropy~\cite{ingle_nonrigid_2014}. \texttt{NR-REACT} yielded lower image gradient entropy than \texttt{NR-Tran} ($p<10^{-4}$). For the selected comparisons among the REACT variants, \texttt{Tran}, and \texttt{NoCo}, both image gradient entropy and u-IEPA for the LAD favored the same methods, with all differences significant ($p<0.05$).

\section{Additional Discussion}\label{supp:sec:discussion}

In this study, simple 3D ellipsoids with manually selected centers and semi-axes were used as binary ROI masks ($\mathbf{W}$ in (5)) to compute the image-quality metric. Refined masks that more selectively target the ROI and exclude stationary image artifacts unrelated to motion, such as bSSFP banding artifacts, may further improve performance.

As described in (1) and (2), the proposed method exploits the shift and rotation properties of the Fourier transform to efficiently apply 3D rigid transformations to MR data. Using an analogous formulation that relates affine transformations in image space to those in $k$-space~\cite{bracewell_affine_1993}, the proposed method could be extended to 3D affine motion estimation.

\clearpage

\begin{table*}[p]%
  \caption{Two-sided Paired t-tests for Selected Comparisons.\label{supp:tab:t-test-entropy}}
  \vspace{6pt}
  {%
  \begin{tabular*}{\textwidth}{@{\extracolsep{\fill}} ll cc @{}}
    \toprule
    \multicolumn{2}{c}{\textbf{Comparisons} (A $\neq$ B)} 
    & \multicolumn{2}{c}{\textbf{Image Gradient Entropy Reduction}} \\
    \cmidrule(lr){1-2} \cmidrule(lr){3-4}
    \textbf{A}      & \textbf{B} 
    & \textbf{Diff} & \textbf{p-value} \\
    \midrule
    Tran       & NoCo           &~0.046 $\pm$ 0.020 & 4.7e-5 \\
    NR-Tran    & Tran           &~0.256 $\pm$ 0.041 & 9.5e-9 \\
    REACT      & Tran           &~0.084 $\pm$ 0.022 & 7.0e-7 \\
    SN-REACT   & Tran           &~0.075 $\pm$ 0.020 & 8.7e-7 \\
    REACT-Tran & Tran           &~0.044 $\pm$ 0.020 & 5.7e-5 \\
    REACT      & REACT-Tran     &~0.040 $\pm$ 0.007 & 1.9e-8 \\
    REACT      & NR-Tran        &-0.172 $\pm$ 0.054 & 3.4e-6 \\
    NR-REACT   & NR-Tran        &~0.051 $\pm$ 0.019 & 1.1e-5 \\
    REACT      & REACT-NoGrp    &~0.009 $\pm$ 0.007 & 2.1e-3 \\
    SN-REACT   & SN-REACT-NoGrp &~0.005 $\pm$ 0.005 & 1.0e-2 \\
    \bottomrule
  \end{tabular*}
  }%
  \begin{tablenotes}
    \item Values represent the mean $\pm$ standard deviation of image gradient entropy reductions (B - A). 
  \end{tablenotes}
\end{table*}

\clearpage
\end{bibunit}

\end{document}